\pdfoutput=1
\documentclass[10pt,conference]{IEEEtran}
\IEEEoverridecommandlockouts
\usepackage{cite}
\usepackage{amsmath,amssymb,amsfonts,amsthm}
\usepackage{graphicx}
\usepackage{textcomp}
\usepackage{xcolor}
\usepackage{bbding}
\usepackage{algorithm}
\usepackage{algorithmicx}
\usepackage{algpseudocode}
\usepackage[mathscr]{eucal}
\usepackage{multirow}
\usepackage{setspace}
\usepackage{color,framed}
\usepackage{xcolor}
\usepackage{mathrsfs}
\usepackage{tabularx}
\usepackage{utfsym}
\usepackage{soul}
\usepackage{subcaption}
\usepackage{float}
\usepackage{url}
\usepackage{enumitem}
\usepackage{flushend}
\usepackage{booktabs}
\usepackage{threeparttable}
\usepackage[colorlinks,bookmarksopen,bookmarksnumbered,citecolor=green]{hyperref}
\usepackage{pifont}
\usepackage{xspace}

\usepackage[most]{tcolorbox}
\definecolor{gray}{RGB}{216,216,216}

\newif\ifshowcomments
\showcommentstrue

\ifshowcomments
\newcommand{\wu}[1]{\mytodored{[rongxin: #1]}}
\else
\newcommand{\wu}[1]{}
\fi

\newcommand{\mytodored}[1]{\textcolor{red}{\ding{46}~{\sf}~#1}}

\newcommand{\code}[1]{{\fontfamily{cmtt}\fontseries{m}\fontshape{n}\selectfont\small{#1}}}

\def\BibTeX{{\rm B\kern-.05em{\sc i\kern-.025em b}\kern-.08em
    T\kern-.1667em\lower.7ex\hbox{E}\kern-.125emX}}
\begin{document}

\title{Improving Java Deserialization Gadget Chain Mining via Overriding-Guided Object Generation
\thanks{$^\ast$Work done during internship at Ant Group.}
\thanks{\textsuperscript{\Envelope}Corresponding authors.}
}

\author{
    \IEEEauthorblockN{Sicong Cao$^{\dag\ast}$, Xiaobing Sun$^\dag$\textsuperscript{\Envelope}, Xiaoxue Wu$^\dag$\textsuperscript{\Envelope}, Lili Bo$^\dag$\textsuperscript{\Envelope}, Bin Li$^\dag$, Rongxin Wu$^\ddag$, \\
    Wei Liu$^\dag$, Biao He$^\S$, Yu Ouyang$^\S$, Jiajia Li$^\S$}
    \IEEEauthorblockA{$^\dag$Yangzhou University\quad $^\ddag$Xiamen University\quad $^\S$Ant Group}
    \IEEEauthorblockA{$^\dag$\{DX120210088, xbsun, xiaoxuewu, lilibo, lb, weiliu\}@yzu.edu.cn, $^\ddag$wurongxin@xmu.edu.cn,}
    \IEEEauthorblockA{$^\S$\{hb187361, yu.oyy, jiajia.lijj\}@antgroup.com}
}

\maketitle

\begin{abstract}
Java (de)serialization is prone to causing security-critical vulnerabilities that attackers can invoke existing methods (gadgets) on the application's classpath to construct a gadget chain to perform malicious behaviors. Several techniques have been proposed to statically identify suspicious gadget chains and dynamically generate injection objects for fuzzing. However, due to their incomplete support for dynamic program features (e.g., Java runtime polymorphism) and ineffective injection object generation for fuzzing, the existing techniques are still far from satisfactory.

In this paper, we first performed an empirical study to investigate the characteristics of Java deserialization vulnerabilities based on our manually collected 86 publicly known gadget chains. The empirical results show that 1) Java deserialization gadgets are usually exploited by abusing runtime polymorphism, which enables attackers to reuse serializable overridden methods; and 2) attackers usually invoke exploitable overridden methods (gadgets) via dynamic binding to generate injection objects for gadget chain construction. Based on our empirical findings, we propose a novel gadget chain mining approach, \emph{GCMiner}, which captures both explicit and implicit method calls to identify more gadget chains, and adopts an overriding-guided object generation approach to generate valid injection objects for fuzzing. The evaluation results show that \emph{GCMiner} significantly outperforms the state-of-the-art techniques, and discovers 56 unique gadget chains that cannot be identified by the baseline approaches.
\end{abstract}


\section{Introduction}
Java serialization \cite{DBLP:journals/toplas/HerlihyL82} enables an application to convert an object to a stream of bytes. By contrast, Java deserialization reconstructs the original object from its serialized byte stream. In spite of the convenience of Java serialization in cross-platform data transmission and persistence storage \cite{SALSA}, deserializing data from untrusted provenance provides an entry point for diverse attacks, including denial of service (DoS) attacks \cite{DBLP:journals/smr/WeiSBCXL21,DBLP:journals/chinaf/SunPZLC19} and remote code execution (RCE) \cite{Svoboda}. In such attacks, an attacker reuses exploitable code fragments (so called \emph{gadgets}) on the application's classpath and joins them together piece by piece (so called \emph{gadget chains}) to facilitate a malicious injection object flowing into the security-sensitive call site (e.g., \code{Method.invoke()}) \cite{DBLP:conf/ccs/DahseKH14}. A deserialization vulnerability disclosed recently, namely \code{Spring4Shell} \cite{Spring4Shell}, allows attackers to send queries to create web shells to servers running the \code{Spring} framework \cite{Spring}, leading to RCE. The impact of the \code{Spring4Shell} vulnerability can be devastating because 60\% of developers use \code{Spring} for their Java applications development. Due to its severity, OWASP (Open Web Application Security Project) lists \emph{Insecure Deserialization} as the top 10 most critical web application security risks in 2017 \cite{OWASP}.

The root cause of deserialization vulnerabilities is that  the control and data flow  can be manipulated by attackers to enable the malicious deserialized objects to \emph{reach} (in terms of control flow) and \emph{affect} (in terms of data flow) the security-sensitive sink \cite{DBLP:conf/ccs/DahseKH14}.  Some techniques \cite{Inspector,Serhybrid} have been proposed to automatically mine exploitable gadget chains. \emph{Gadget Inspector} \cite{Inspector} performs static taint analysis \cite{DBLP:journals/ese/LuoPPBPMBHM22,DBLP:conf/icse/BenzKLBBZ20} to track inter-procedural data flows, and applies the Breadth-First Search (BFS) to identify  gadget chains  from deserialization entry points (sources) to security-sensitive call sites (sinks). Considering that such a purely static solution may suffer from precision issues and requires manual inspection of the reports, SerHybrid \cite{Serhybrid} adopts a hybrid analysis solution, which constructs the heap access paths to find source objects that affect security-sensitive call sites and utilizes fuzzing \cite{fuzz,fuzz1} to generate actual injection objects, to verify whether the sinks are reachable.

Nevertheless, \emph{SerHybrid} has limited effectiveness due to two reasons. First, \emph{SerHybrid} performs points-to analysis \cite{DBLP:conf/pldi/Ruf95} to identify source-to-sink method execution paths. However, due to the dynamic features (e.g., runtime polymorphism \cite{sound2}) of Java language, any available overridden method (gadget) on the application's classpath may be exploited to construct gadget chains, resulting in high false negatives. Second, \emph{SerHybrid} generates injection objects based on heap access paths for gadget chain verification. This heap access path reflects the taint propagation flow from a source object to the security-sensitive call site. However, due to the unawareness of hard constraints (requiring dynamically modifying the properties of an injection object to trigger the target gadget chain) introduced by certain gadgets, such generated injection objects may be semantically invalid. Hence, fuzzing solutions blind to the structure of a given gadget chain will get stuck in the initial fuzzing stage.

In this paper, we first performed an empirical study to investigate the characteristics of Java deserialization vulnerabilities. In particular, we focused on answering the following two questions: 1) how Java deserialization gadgets are exploited; and 2) how gadget chains are constructed. We manually constructed a real-world Java deserialization vulnerability benchmark, which consists of the well-known \emph{ysoserial} \cite{ysoserial} repository and 18 popular Java applications. In total, 86 (52 out of which are new) publicly exploitable gadget chains are included in our benchmark. From the empirical results, we found that 1) Java deserialization gadgets are usually exploited by abusing advanced language features (e.g., runtime polymorphism), which enables attackers to reuse serializable overridden methods on the application's classpath; and 2) attackers usually invoke exploitable overridden methods (gadgets) via dynamic binding \cite{DBLP:journals/jss/GantenbeinJ88} to generate injection objects for gadget chain construction.

Based on our empirical findings, we propose a novel \underline{G}adget \underline{C}hain \underline{Miner}, \emph{GCMiner}, which considers dynamic program features (Java runtime polymorphism) to identify more exploitable gadget chains and generates valid injection objects through dynamic binding for fuzzing. \emph{GCMiner} performs static analysis to construct the \emph{Deserialization-Aware Call Graph} (DA-CG) to model both explicit and implicit (method overriding) method calls to identify more gadget chains. To verify whether a statically identified gadget chain is exploitable, \emph{GCMiner} adopts an overriding-guided object generation approach to generate exploitable injection objects for fuzzing. Using the overriding relations between methods as a guidance for injection object generation can effectively help the fuzzer to be aware of the object structures that reflect the target gadget chain. Gadget chains which receive an injection object reachable to their security-sensitive call sites will be outputted as exploitable chains.

To evaluate the effectiveness, we compared \emph{GCMiner} with two state-of-the-art gadget chain mining tools, \emph{Gadget Inspector} \cite{Inspector} and \emph{Serhybrid} \cite{Serhybrid}. Our experimental results demonstrate that \emph{GCMiner} significantly outperforms the baselines. 

In summary, this paper makes the following contributions:
\begin{itemize}[leftmargin=1em]
    \item \textbf{An empirical study:} We performed an empirical study on 86 exploitable gadget chains from several famous open-source Java projects. Our findings show that dynamic features of Java language are abused to construct gadget chains by generating injection objects through dynamically binding exploitable gadgets.
    \item \textbf{A novel gadget chain mining approach:} We propose a gadget chain mining approach that identifies more exploitable gadgets by considering deserialization-related dynamic features of Java language, and generates injection objects through dynamically binding overridden methods to verify statically reported gadget chains. We open-sourced \emph{GCMiner} and the benchmark to facilitate further research\footnote{https://github.com/GCMiner/GCMiner}.
    \item \textbf{Technique evaluation:} We performed comprehensive experiments to evaluate the effectiveness of \emph{GCMiner}. The experimental results reveal that \emph{GCMiner} discovers 56 unique gadget chains missed by state-of-the-art baselines.
\end{itemize}

\section{Background and Motivation}\label{Motivation}

\subsection{Terminology}
In this section, we introduce several basic concepts  which will be used  in this paper.

\noindent\textbf{Java deserialization vulnerability.}
A Java deserialization vulnerability is a security bug that can be exploited when the Java application deserializes untrusted data. Attackers could inject crafted objects into deserialization-related methods (e.g., \code{readObject()}), which pass malicious commands to a security-sensitive call site, resulting in diverse attacks like RCE.

\noindent\textbf{Gadget and gadget chain.}
To facilitate an injection object reaching the security-sensitive call site, attackers should reuse a series of exploitable methods in memory to manipulate the deserialization process to achieve their desired malicious behaviors. Such a sequence of method calls is called a \emph{gadget chain}, and each method of this chain is called a \emph{gadget} \cite{DBLP:conf/ccs/DahseKH14}. The presence of a gadget chain on the application's classpath is one of the necessary conditions to carry out deserialization attacks.

\noindent\textbf{Magic method and security-sensitive call site.}
The security-risk of a gadget chain comes from its first gadget/method (source) that can be invoked \emph{automatically} during object deserialization. Such a self-executing method is called a \emph{magic method}. These magic methods can be exploited by attackers to bypass existing security defenses to deserialize their crafted injection objects. Correspondingly, the last gadget/method (sink) which performs the malicious command carried by the injection object is called a \emph{security-sensitive call site}.

\noindent\textbf{Property-oriented programming.}
To exploit a Java deserialization vulnerability, attackers have to instantiate an object of attacker-controlled type and modify its properties to trigger the execution of an exploitable gadget chain. Such a technique used in constructing this injection object is called \emph{Property-Oriented Programming} (POP) \cite{POP}. POP allows an attacker to manipulate the data- and control-flow of the victim application, thereby exploiting existing gadgets on the application's classpath for deserialization attacks.

\subsection{Motivating Example}

\begin{figure}[t]
\centering
\begin{subfigure}{\linewidth}
\centering
\includegraphics[width=\linewidth]{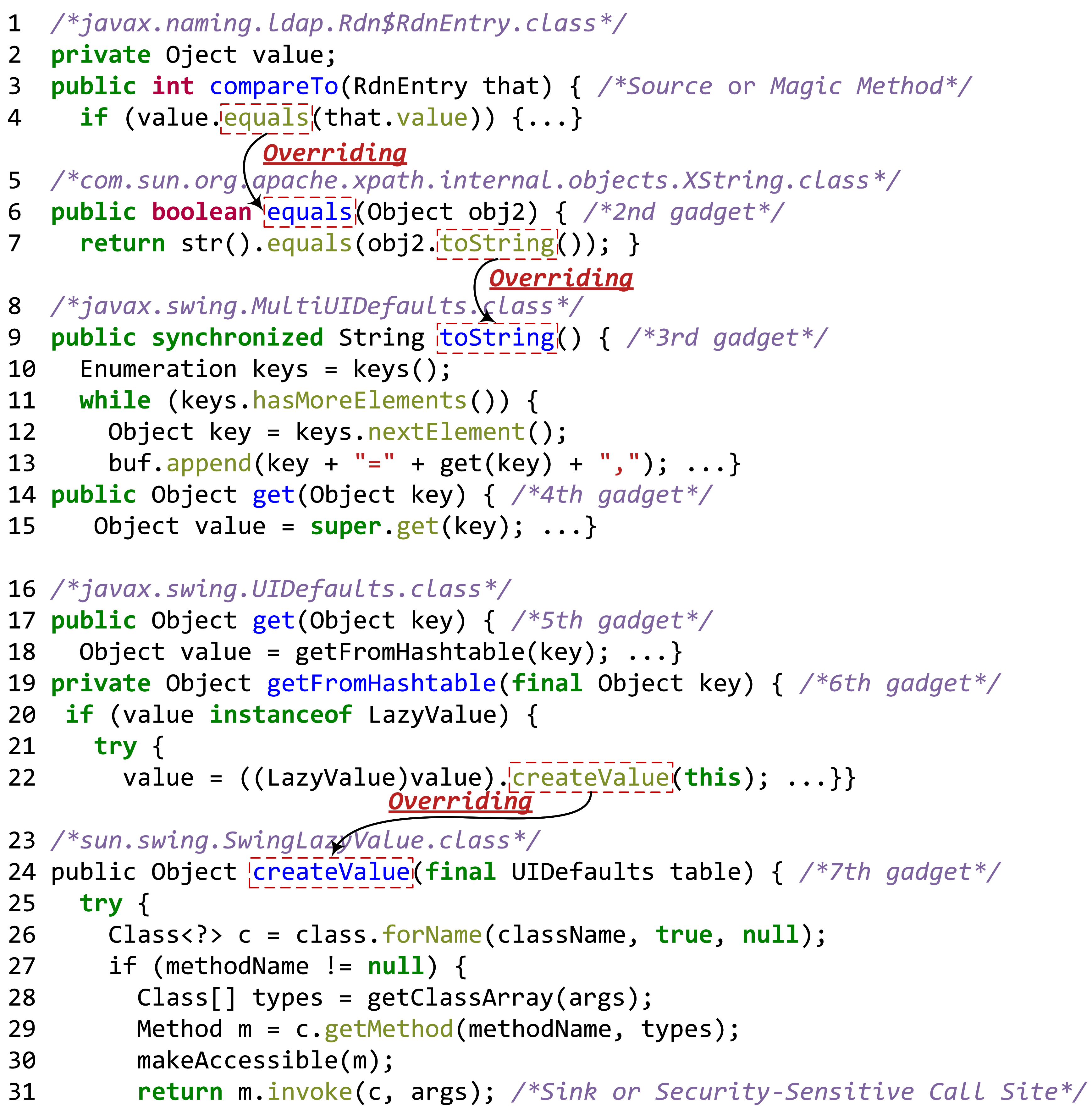}
\caption{\label{motivating example}A simplified code snippet with deserialization vulnerability.}
\quad
\end{subfigure}
\begin{subfigure}{\linewidth}
\centering
  \includegraphics[width=.9\linewidth]{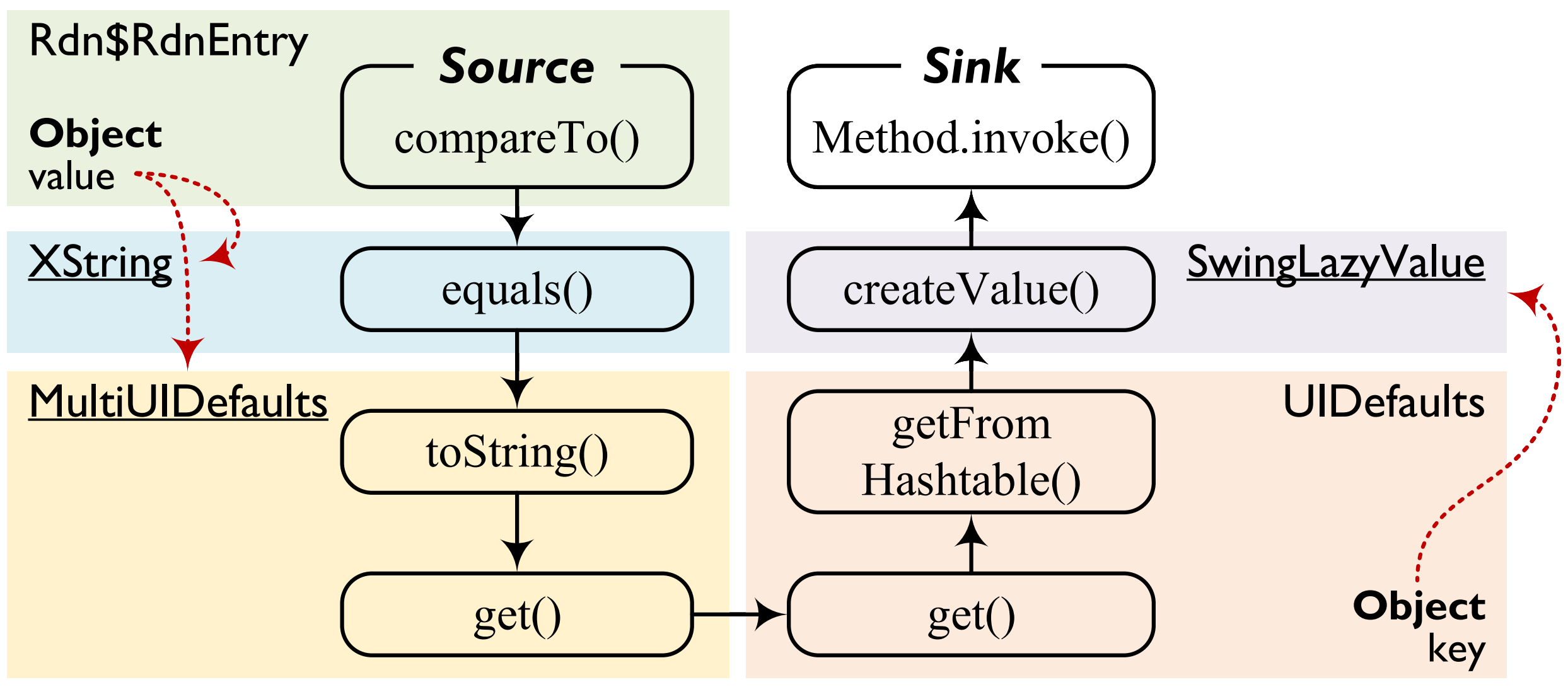}
\caption{\label{popchain}The stack trace of the gadget chain in Figure \ref{motivating example}.}
\end{subfigure}
\caption{\label{CodeChain}An example Java deserialization vulnerability and the gadget chain.}
\end{figure}

To better clarify the above terminology and illustrate the motivation of our approach, we take a real-world Java deserialization vulnerability (CVE-2021-21346\footnote{https://x-stream.github.io/CVE-2021-21346.html}) in \code{XStream} \cite{XStream} as an example. \code{XStream} is a popular Java deserialization library to serialize objects to XML and back again. 

Figure \ref{motivating example} presents the simplified code snippet of the target gadget chain in CVE-2021-21346. The serializable class \code{javax.naming.ldap.Rdn} (line 1) invokes the magic method \code{compareTo()} (line 3) automatically during object deserialization. Such a magic method can be regarded as the \emph{source} or \emph{entry point} that allows attackers to inject malicious objects. If a program performs object deserialization without additional security check, an arbitrary method (e.g., \code{Runtime.exec(command)} \cite{10.1145/3554732}) specified in a string \code{className} (line 26) carried by this malicious injection object will be invoked (line 31), leading to RCE.

To exploit this vulnerability, an attacker chains existing gadgets on the application's classpath to enable the malicious injection object to flow from the source to the sink. For example, to invoke the second gadget \code{equals()} at line 6, an attacker instantiates the class \code{Xstring} (line 5) and assigns this instance to the injection object \code{Rdn\$RdnEntry}'s property \code{value} (line 2) by POP. This dynamic method call takes advantage of Java polymorphism. Similarly, the third gadget (\code{toString()} at line 9) and seventh gadget (\code{createValue()} at line 24) should also be dynamically invoked to facilitate the propagation of the injection object. Note that the fourth gadget (\code{get()} at line 14) to sixth gadget (\code{getFromHashtable} at line 19) can be directly invoked during deserialization. Figure \ref{popchain} depicts the corresponding stack trace of the gadget chain. In general, constructing such an exploitable gadget chain requires 1) modifying the injection object’s properties to trigger the execution of an exploitable gadget chain, and 2) assigning proper values to reach the security-sensitive call site.

\begin{tcolorbox}[left=1pt,right=1pt,top=1pt,bottom=1pt,boxrule=0.7pt,enhanced,drop fuzzy shadow]
\textbf{[Observation-1] }\textit{Dynamic program features (e.g., Java runtime polymorphism) can be abused by attackers to implement insecure deserialization paths.}
\end{tcolorbox}

As shown in Figure \ref{motivating example}, since class \code{XString} to which the gadget \code{equals()} (line 6) belongs inherits the superclass \code{Object}, the tainted property \code{value} of an injection object \code{RdnEntry} can continue to propagate through the call statement \code{value.equals()} (line 4). Taking such a dynamic runtime behavior into account can effectively identify attacker-controlled gadgets on the application's classpath. A straightforward solution is to build Call Graphs (CGs) \cite{CG} to perform inter-procedural analysis, as existing gadget chain mining tools \cite{Inspector,Serhybrid} do. However, these statically constructed CGs either ignore these serialization-related dynamic features or handle them partially \cite{DBLP:conf/icse/Sui0TF20,sound2}, resulting in unsound results.

Therefore, for gadget chain identification, the consideration of dynamic program features may help avoid false negatives.

\begin{tcolorbox}[left=1pt,right=1pt,top=1pt,bottom=1pt,boxrule=0.7pt,enhanced,drop fuzzy shadow]
\textbf{[Observation-2] }\textit{Constructing an exploitable gadget chain requires dynamically invoking a series of overridden methods to enable the tainted objects  to pass into the dangerous sink.}
\end{tcolorbox}

Due to the imprecision of static analysis \cite{DBLP:conf/issta/NachtigallSB22}, most statically reported gadget chains cannot be exploited in practice, resulting in a high \emph{false-positive rate}. To alleviate this problem, generating exploitable injection objects to verify the exploitability of suspicious gadget chains is an effective solution \cite{thanassis2011aeg,AEG1,AEG2}. However, generating injection objects that follow the execution trace of gadget chains to be verified is challenging because of the nested class hierarchy, i.e., the properties of an injection object need to be modified to reach the dangerous sink. For instance, to construct the deserialization gadget chain in Figure \ref{motivating example}, an attacker needs to invoke attacker-controlled overridden methods (e.g., \code{equals()} at line 6) on the application's classpath multiple times to implement a customized deserialization routine. This dynamic deserialization behavior can be realized through POP, and is implemented by determining the method to invoke at runtime (i.e., dynamic binding). In this motivating example, the purpose of the first two dynamic bindings is to invoke \code{XString.equals()} (line 6) and \code{MultiUIDefaults.toString()} (line 9) respectively to facilitate the propagation of the tainted property \code{value} (line 2), while the third is to ensure that the injection object can reach the security-sensitive call site \code{Method.invoke()} (line 31) to execute malicious commands. However, the heap access path adopted by \emph{SerHybrid} fails to infer these implicit control flows,
making it hard to reach dangerous sinks.
%resulting in ineffective (hard to reach dangerous sinks) fuzzing.

Therefore, for gadget chain verification, generating injection objects to trigger given gadget chains via dynamic binding may improve the effectiveness of fuzzing.

\section{Empirical Study}\label{Empirical Study}

\subsection{Experiment Setup}\label{Experiment Setup}

Inspired by the above two observations, we performed an empirical study to investigate the characteristics of Java deserialization vulnerabilities. Particularly, we aim to answer the following two research questions:

\begin{itemize}[leftmargin=1em]
    \item \textbf{RQ1: }How are Java deserialization gadgets exploited?
    \item \textbf{RQ2: }How are gadget chains constructed?
\end{itemize}

The answers to these questions provide empirical foundations on 1) whether dynamic program features are abused by attackers to implement insecure deserialization; and 2) whether dynamically invoking overridden methods on the application's classpath are widely exploited to construct gadget chains.

\noindent\textbf{Data collection. }We chose \emph{ysoserial} \cite{ysoserial} repository, a famous project that provides 34 Java payloads with corresponding gadget chains exploited in publicly known deserialization attacks, as part of our dataset. Considering the scarcity of gadget chains available for analysis, we \emph{manually} collected public Java deserialization gadget chains from well-known vulnerability disclosure platforms such as National Vulnerability Database (NVD) \cite{NVD}, Common Vulnerabilities and Exposures (CVE) \cite{CVE}, Exploit Database (Exploit-DB) \cite{Exploit-DB}, etc. We selected target applications that satisfied the following criteria. First, they are Java open-source projects since the characteristics of deserialization vulnerabilities might vary among different programming languages \cite{shcherbakov2021serialdetector}. Second, they support deserialization operations and have been reported to have the risk of being exploited so that gadget chains we mined are available. Third, they contain sufficient information (e.g., \emph{Proof-of-Concept} (POC) and affected versions) to verify the authenticity of gadget chains. 

\begin{table}[t]\footnotesize
  \centering
  %\fontsize{7.5}{8.5}\selectfont
  \renewcommand\arraystretch{1.1}
  \renewcommand\tabcolsep{10pt}
  \caption{Benchmark information.}
    \begin{tabular}{|c|l|c|c|}
    \hline
    \textbf{Library}     & \textbf{Affected Application} & \textbf{\#Chain} & \textbf{Type} \cr
    \hline
    \hline
     - & ysoserial  &  34  & - \cr
    \hline
    \hline
    \multirow{4}{*}{YAML} & JBoss RESTEasy & 1&  \multirow{4}{*}{RCE} \cr
    ~ & Apache Camel  & 2  & ~ \cr
    ~ &Apache Brooklyn  & 1  &~\cr
    ~ &Apache XBean  & 1  & ~\cr
    \hline
    \multirow{2}{*}{JDK} & Shiro & 3 & JNDIi\cr
    ~ & Pippo  & 2 & RCE\cr
    \hline
    \multirow{2}{*}{BlazeDS} & Adobe Coldfusion & 2& \multirow{2}{*}{RCE}\cr
    ~ & VMWare VCenter  & 1 & ~  \cr
    \hline
    Red5  & Red5 &  1 & RCE \cr
    \hline
    Hessian & Hessian & 5 &   RCE\cr
    \hline    
    XStream & XStream & 14 &   RCE SRA\cr
    \hline
    \multirow{7}{*}{Others} & Commons Collections & 3   & RCE\cr
    ~ & Dubbo & 2 & RCE\cr
    ~ & WebLogic & 5 & RCE JNDIi\cr
    ~ & Emissary & 3 & SSRF \cr
    ~ & Jenkins & 2&RCE \cr
    ~ & Apache OFBiz & 3 &RCE\cr    
    ~ & Spring & 1  & JNDIi \cr
    \hline
    \multicolumn{2}{|c|}{\textbf{Total}} &  \textbf{86} &-\cr\cline{1-2}
    \hline
    \end{tabular}
    \label{data}
\end{table}

Table \ref{data} shows the details of the benchmark. Columns ``Library" and ``Affected Application" present the deserialization libraries that cause the vulnerabilities and corresponding affected applications. Note that due to the re-implementation of deserialization operations (i.e., not relying on any deserialization library), column ``Library" of some applications is labeled as ``Others". Column ``\#Chain" represents the number of collected gadget chains. Column ``Type" presents different vulnerability types in each application, including \emph{Remote Code Execution} (RCE), \emph{JDNI Injection} (JNDIi), \emph{System Resource Access} (SRA), and \emph{Server-Side Request Forgery} (SSRF). In total, we collected 86  deserialization gadget chains covering 18 Java applications, 52 out of which are not included in \emph{ysoserial} repository.

\subsection{Exploitation of Java Deserialization Gadgets (RQ1)}\label{RQ1}

An exploitable gadget chain requires: 1) a \emph{magic method} (source or the first gadget) deserializing untrusted data that can be injected by attackers; 2) a \emph{security-sensitive call site} (sink or the last gadget) that ultimately executes a dangerous operation; and 3) a series of \emph{gadgets} facilitating the propagation of injection objects \cite{Munoz}. Hence, we first investigated exploitable magic methods and security-sensitive call sites in our benchmark. They are listed as follows.

\begin{itemize}[leftmargin=1em]
    \item \textbf{Magic methods: }\code{\sethlcolor{gray}{\hl{hashCode}}}, \code{compareTo}, \code{toString}, \code{get}, \code{put}, \code{\sethlcolor{gray}{\hl{compare}}}, \code{\sethlcolor{gray}{\hl{readObject}}}, \code{readExternal}, \code{readResolve}, \code{\sethlcolor{gray}{\hl{finalize}}}, \code{\sethlcolor{gray}{\hl{equals}}}
    \item \textbf{Security-Sensitive Call Sites.}
    
    \textbf{- \emph{Remote Code Execution} (RCE): }\code{getDeclaredMethod}, \code{getConstructor}, \code{\sethlcolor{gray}{\hl{exec}}}, \code{\sethlcolor{gray}{\hl{getMethod}}}, \code{loadClass}, \code{start}, \code{findClass}, \code{\sethlcolor{gray}{\hl{invoke}}}, \code{\sethlcolor{gray}{\hl{forName}}}, \code{\sethlcolor{gray}{\hl{newInstance}}}, \code{defineClass}, \code{<init>}, \code{\sethlcolor{gray}{\hl{exit}}}

    \textbf{- \emph{JDNI Injection} (JNDIi): }\code{getConnection}, \code{connect}, \code{lookup}, \code{getObjectInstance}, \code{do\_lookup}
    
    \textbf{- \emph{System Resource Access} (SRA): }\code{\sethlcolor{gray}{\hl{newBufferedReader}}}, \code{\sethlcolor{gray}{\hl{newBufferedWriter}}},
    \code{delete}, \code{\sethlcolor{gray}{\hl{newInputStream}}}, \code{\sethlcolor{gray}{\hl{newOutputStream}}}
    
    \textbf{- \emph{Server-Side Request Forgery} (SSRF): }\code{openConnection}, \code{\sethlcolor{gray}{\hl{openStream}}}
\end{itemize}

In total, a set of 11 magic methods and 25 security-sensitive call sites are found in our benchmark. It is worth noting that only five magic methods and 11 security-sensitive call sites (highlighted in gray) are included in previous works \cite{Inspector,Serhybrid}. The direct consequence of missing these exploitable magic methods (e.g., \code{compareTo()} in Figure \ref{CodeChain}) and security-sensitive call sites is that some exploitable gadget chains cannot be identified.

\begin{figure}[t]
  \centering 
  \includegraphics[width=.45\linewidth]{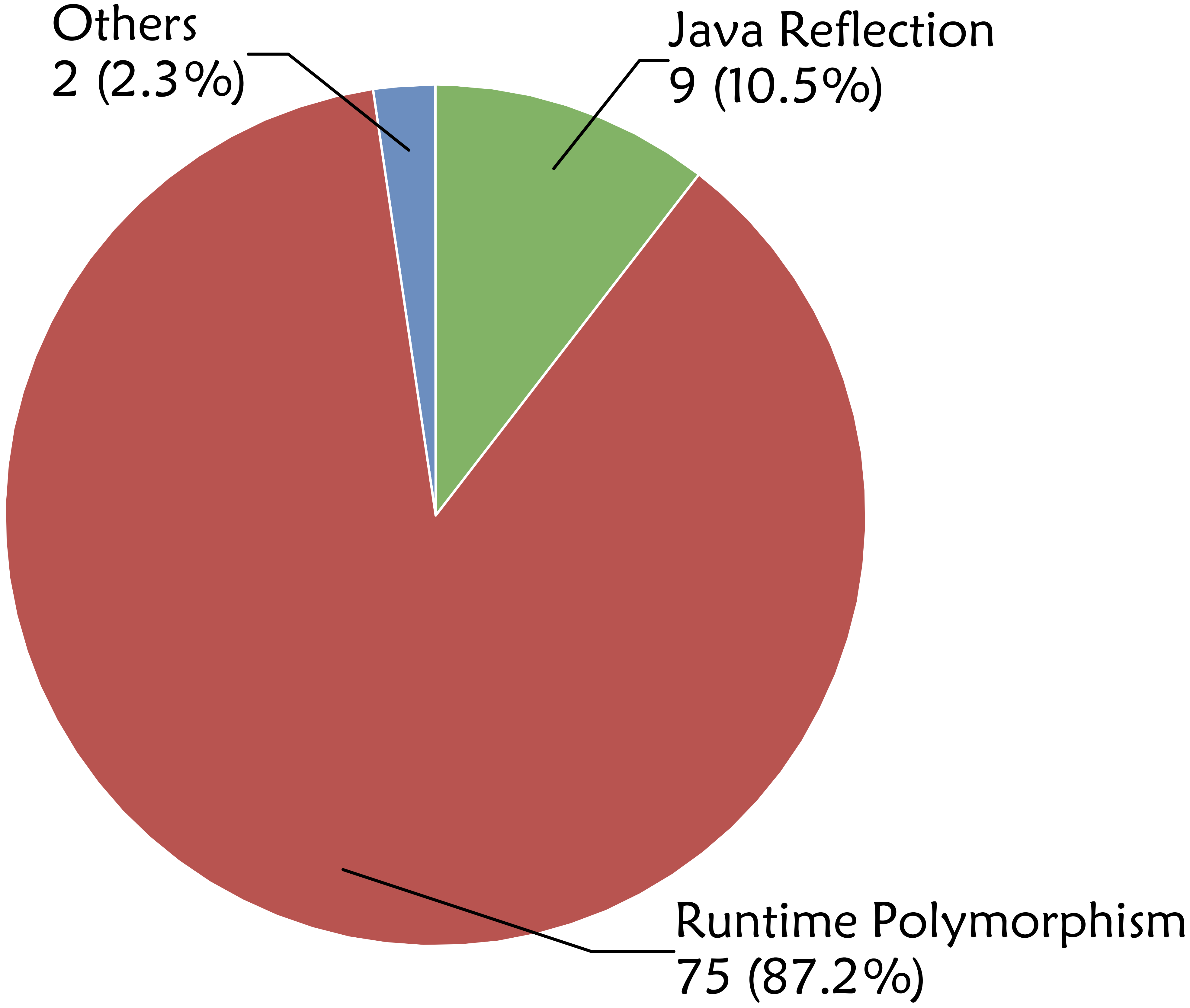}
\caption{Ways of exploiting available gadgets.}
\label{Pie}
\end{figure}

In addition, we further investigated how these gadgets were exploited. In particular, we focused on dynamic program features of Java language that may be abused by attackers. As shown in Figure \ref{Pie}, 75 out of 86 (87.2\%) gadget chains exploit available gadgets by abusing Java runtime polymorphism to invoke overridden methods on the application's classpath. As described before, attackers can exploit these gadgets to pass malicious injection objects. In the remaining cases, only nine (10.5\%) gadget chains rely on Java reflection to exploit gadgets.

%Several gadget chains constructed by newly added magic methods or security-sensitive call sites are shown in Table \ref{source}. However, only a part of these sources and sinks are considered by existing tools. For example, only \texttt{hashCode} and \texttt{compare} are considered as sources in \emph{Serhybrid} \cite{Serhybrid}. The direct consequence of missing these exploitable magic methods and security-sensitive call sites is that quite a lot of gadget chains cannot be identified. By contrast, as the experimental results shown in our evaluation (Section \ref{Ablation}), with our new set of sources and sinks, the mining performance of each tool will be improved. Hence, the new set of magic methods and security-sensitive call sites can help identify more potential gadget chains.

\begin{tcolorbox}[left=1pt,right=1pt,top=1pt,bottom=1pt,boxrule=0.7pt,enhanced,drop fuzzy shadow]
\textbf{[Finding-1] }\textit{Java deserialization gadgets are commonly exploited by abusing advanced language features (e.g., runtime polymorphism), which enables attackers to reuse serializable overridden methods on the application's classpath.}
\end{tcolorbox}

\subsection{Construction of Gadget Chains (RQ2)}\label{RQ2}

\begin{figure}
  \centering 
  \includegraphics[width=0.45\linewidth]{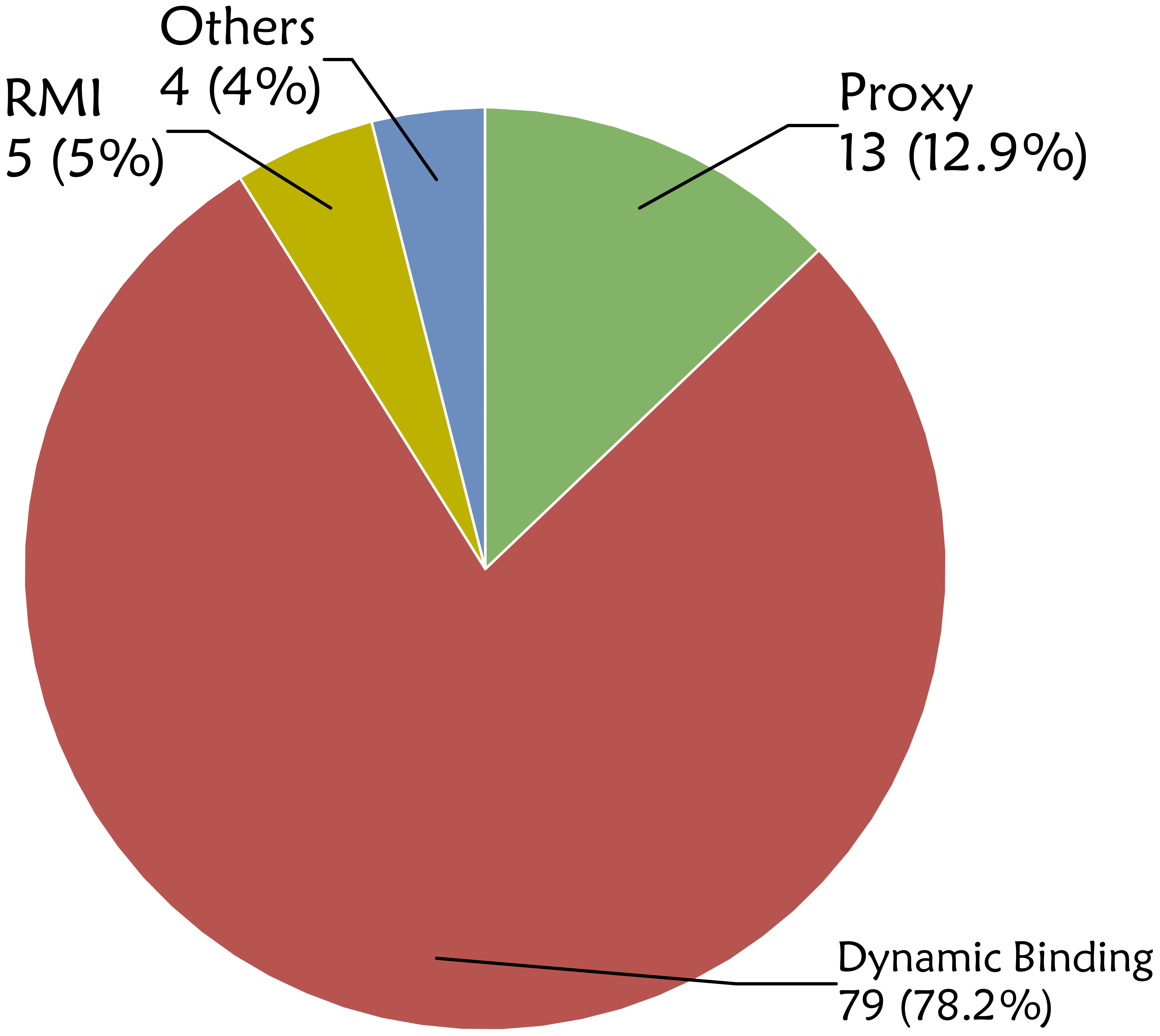}
\caption{Ways of gadget chain construction.}
\label{Column}
\end{figure}

Figure \ref{Column} shows the distribution of different ways for gadget chain construction. The results show that 79 out of 86 (91.9\%) known gadget chains leverage dynamic binding (\emph{at least} once) to modify the properties of injection objects to invoke overridden methods. Besides, in some cases, diverse ways (e.g., remote method invocation (RMI) \cite{RMI} and dynamic proxy \cite{DBLP:conf/issta/FourtounisKS18}) are mixed to trigger the execution of gadget chains\footnote{Note that a gadget chain constructed by multiple techniques will be repeatedly counted in Figure \ref{Column}.}. For example, to exploit a known gadget chain \code{CommonsCollections1} \cite{CommonsCollections1} of \emph{ysoserial}, attackers have to combine dynamic proxy and dynamic binding to invoke exploitable gadgets to reach the security-sensitive call site.

%In this way, we can dynamically construct valid injection objects which can trigger the security-sensitive call sites for automatic verification. Besides, the results of our ablation study on the impact of alias-guided object generation (Section \ref{Ablation}) further demonstrate that more sink-reachable injection objects can be generated by leveraging overriding methods (i.e., alias relations between methods on the classpath) to guide the dynamic binding. For the remaining 13 gadget chains, dynamic binding is not necessary for gadget chain construction because dangerous sinks (e.g., \texttt{invoke}) can be directly invoked once malicious objects are received by native or re-implemented magic methods (e.g., \texttt{readObject}). 

%For example, in \emph{ysoserial} repository, there is a short gadget chain (just three methods are used for constructing a exploitable gadget chain) which is labeled as \emph{URLDNS} \cite{URLDNS}. It can be leveraged to perform a \emph{JNDI Injection} attack without any additional processing like dynamic proxy and reflection once it receives an unsafe serialized object.

\begin{tcolorbox}[left=1pt,right=1pt,top=1pt,bottom=1pt,boxrule=0.7pt,enhanced,drop fuzzy shadow]
\textbf{[Finding-2] }\textit{To construct exploitable gadget chains, attackers usually invoke exploitable overridden methods (gadgets) via dynamic binding to generate injection objects, which facilitate the malicious data flowing into dangerous sinks.}
\end{tcolorbox}

\section{Methodology}\label{Methodology}

\begin{figure}
  \centering
  \includegraphics[width=0.75\linewidth]{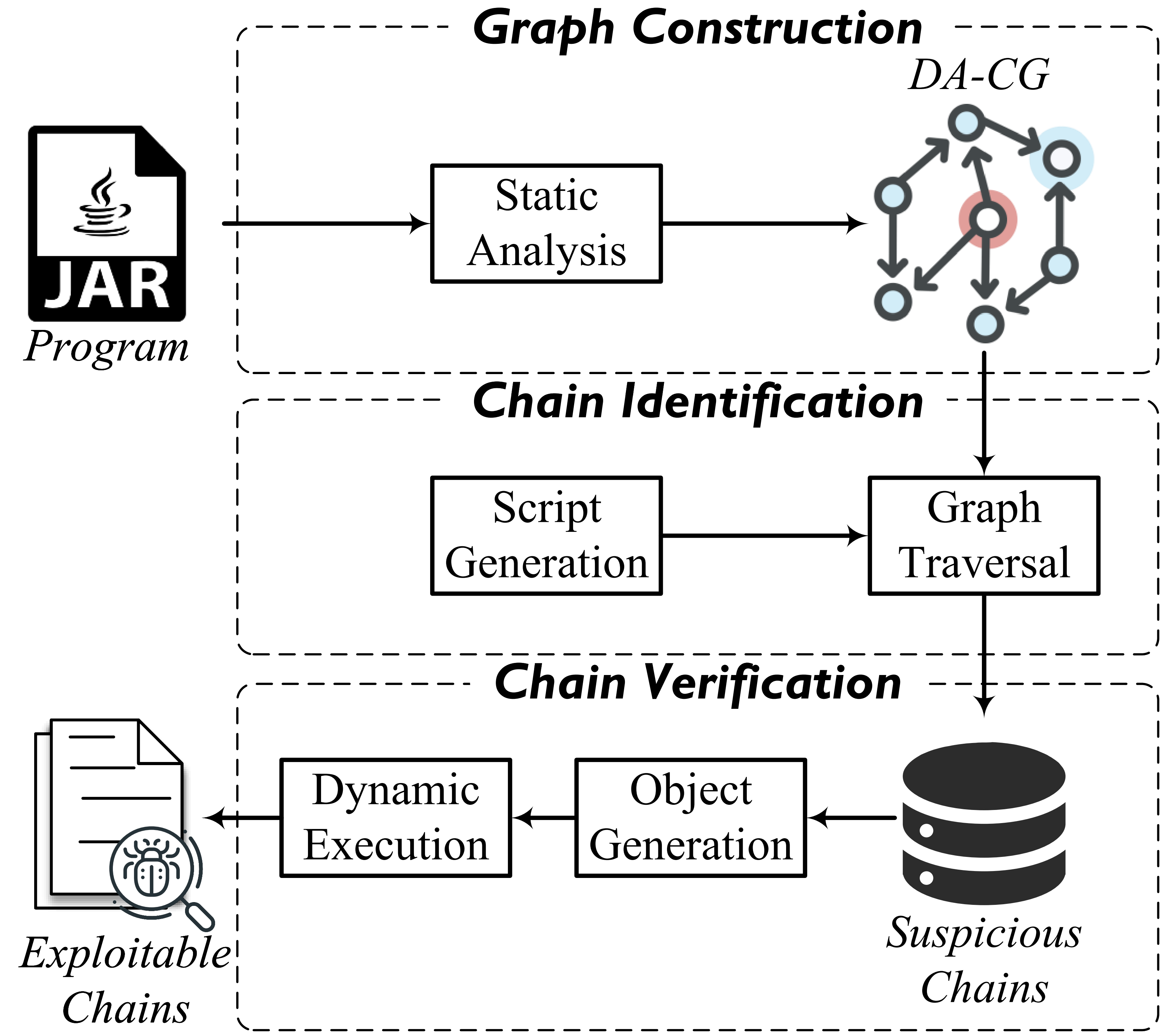}
\caption{Framework of \emph{GCMiner}.}
\label{overview}
\end{figure}

Based on our empirical findings, we propose \emph{GCMiner}, a novel gadget chain mining approach which takes dynamic program features (Java runtime polymorphism) into account to mine implicit method calls for gadget chain identification, and generates valid injection objects through dynamic binding for fuzzing.

Figure \ref{overview} shows the framework of \emph{GCMiner}, which contains three modules: \emph{Graph Construction}, \emph{Chain Identification}, and \emph{Chain Verification}. More specifically, \emph{GCMiner} takes a target Java application as the input, and constructs the \emph{Deserialization-Aware Call Graph} (DA-CG) through static analysis to model both explicit and implicit method calls (Section \ref{Graph Construction}). Then, \emph{GCMiner} stores the DA-CG into the graph database and searches for suspicious gadget chains through graph traversal (Section \ref{Chains Identification}). Finally, to verify whether a statically identified gadget chain is exploitable, \emph{GCMiner} adopts an overriding-guided object generation approach to generate exploitable injection objects for fuzzing (Section \ref{Chains Verification}). The gadget chain which receives an injection object reachable to a security-sensitive call site will be confirmed as exploitable.

\subsection{Graph Construction}\label{Graph Construction}
\emph{GCMiner} first performs static analysis to generate the \emph{Call Graph} (CG) \cite{CG} to capture explicit method calls. Considering that statically constructed CGs may miss some exploitable methods due to their incomplete support for dynamic program features (as discussed in Section \ref{RQ1}), we add additional overriding relations through \emph{Class Hierarchy Analysis} (CHA) \cite{CHA} to construct a \emph{Deserialization-Aware Call Graph} (DA-CG) to identify implicit method calls. The DA-CG is represented as a directed graph, where methods in each class are graph nodes, and two types of directed relations (i.e., method call and method overriding) between methods are recorded as edges. The introduction of overriding relations contributes to model dynamic behaviors of programs and thereby capturing more exploitable gadgets missed by pure CGs.

It is noteworthy that blindly considering all possible overridden methods on the application's classpath will introduce a large number of false positives. A typical example is \code{toString()}, one of the common magic methods. For a large Java application, a great number of classes re-implement \code{toString()} to transform the reference to an object to a user-readable form. To this end, for applications/libraries (e.g., \code{Apache Commons Collections}) which do not re-implement their own deserialization operations, we ignore methods whose classes that do not support deserialization operations (i.e., not implementing serialization interfaces like \code{Serializable}) to focus on deserialization-related method calls because these methods cannot be invoked during object deserialization. For those applications/libraries (e.g., \code{XStream}) which re-implement their own deserialization operations, we still consider all possible methods.

\begin{figure}
  \centering
  \includegraphics[width=.9\linewidth]{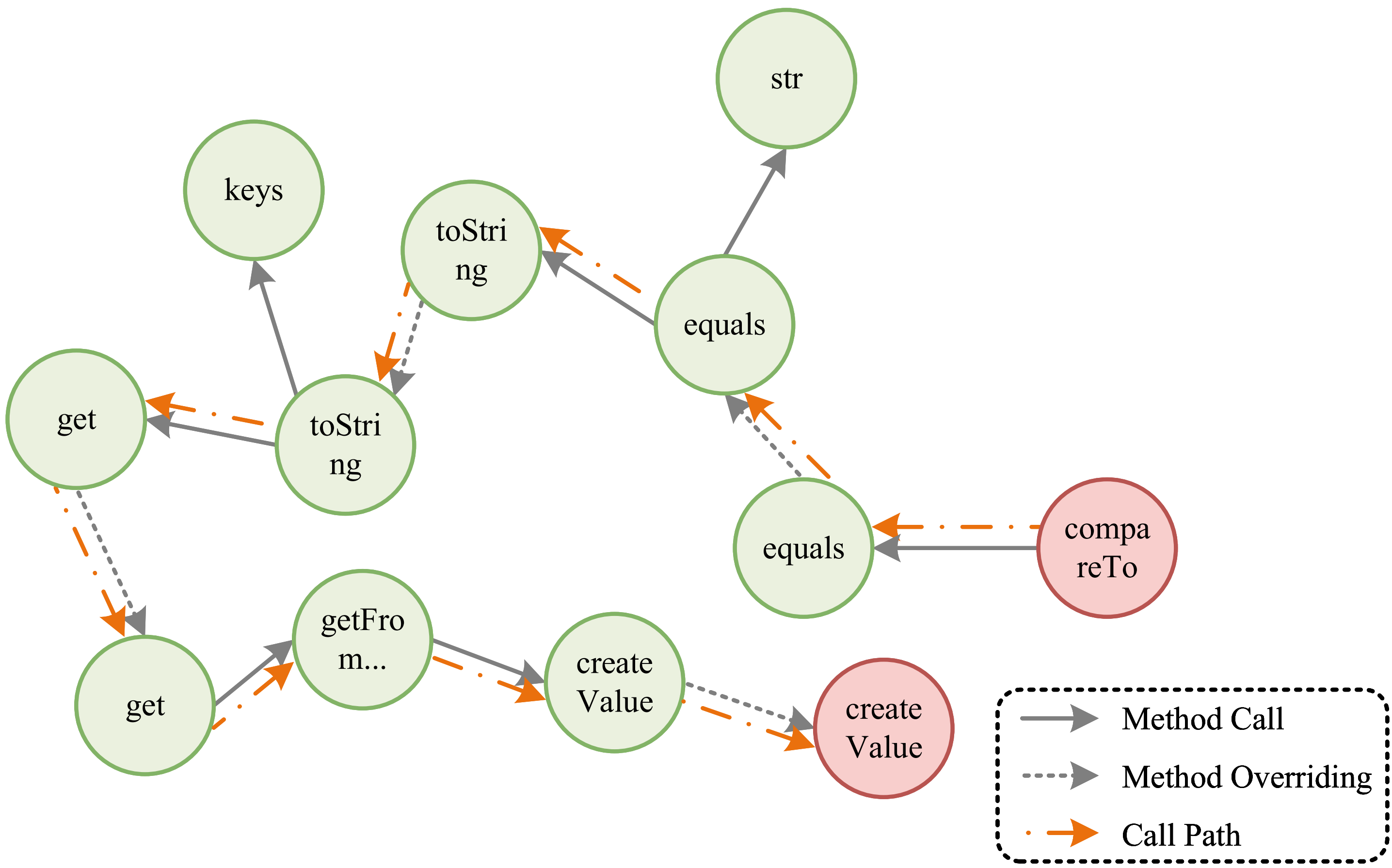}
\caption{Partial DA-CG for our motivating example.}
\label{graph}
\end{figure}

\noindent\textbf{Example.} Figure \ref{graph} shows a part of our constructed DA-CG for the motivating example in Figure \ref{CodeChain}. We can observe that the DA-CG consists of several method nodes (highlighted in green and red) and two types of edges (method call and method overriding). The complete call path (orange shaded) of the gadget chain in the motivating example is clear, which starts from the magic method \code{CompareTo()} to the security-sensitive call site \code{Createvalue()}. We can observe that, owing to our additional overriding relations, exploitable gadgets (e.g., \code{XString.equals()} at line 6 and \code{MultiUIDefaults.toString()} at line 9 in Figure \ref{motivating example}) can also be identified.

% A blue class node \texttt{Multiuidefaults} points to one of the two \texttt{toString} methods through Has edge, which indicates that this \texttt{toString} method is located in class \texttt{Multiuidefaults}. The other \texttt{toString} method belongs to class \texttt{Object}, which is the super class of all classes. Therefore, we can infer that there is Alias relationship between these two methods and connect them with the Alias edge.

% Due to the inheritance characteristics of Java, when the subclass has the same method as its super class and a variable in the super class calls an object of the subclass, which method to call is determined by the real type of the object. Therefore, we introduce the Alias edge which connects two methods with the same name to capture the alias relation and apply Class nodes to retain information of these methods' class. Figure \ref{graph} shows that there are two method nodes with the same name \texttt{toString}, which are connected by Alias edge.  At the same time, a blue class node \texttt{Multiuidefaults} points to one of the two toString methods through Has edge, which indicates that this \texttt{toString} method is located in class \texttt{Multiuidefaults}.

\subsection{Chain Identification}\label{Chains Identification}

After graph construction, we store the DA-CG into the graph database and search for suspicious gadget chains through customized query scripts.

Specifically, we adopt \emph{Cypher} \cite{Cypher}, a declarative language for graph data retrieval \cite{DBLP:conf/sigsoft/Cheng0BW22}, to design query scripts for suspicious gadget chain identification. The script mainly includes three components, i.e., start nodes (sources), end nodes (sinks), and path constraints. We use 11 magic methods and 25 security-sensitive call sites found in our empirical study as sources and sinks to limit the retrieval scope. For path constraints, we consider two types of edges (i.e., method call and method overriding) in our DA-CG. Note that, although the overriding relation does not indicate the actual method call, it provides a hint that these overridden methods can be exploited as gadgets to propagate malicious injection objects. Thus, to avoid missing any suspicious gadget chain, we search for all gadget chains between sources and sinks as candidates for verification.

\begin{figure}[t]
  \centering
  \includegraphics[width=\linewidth]{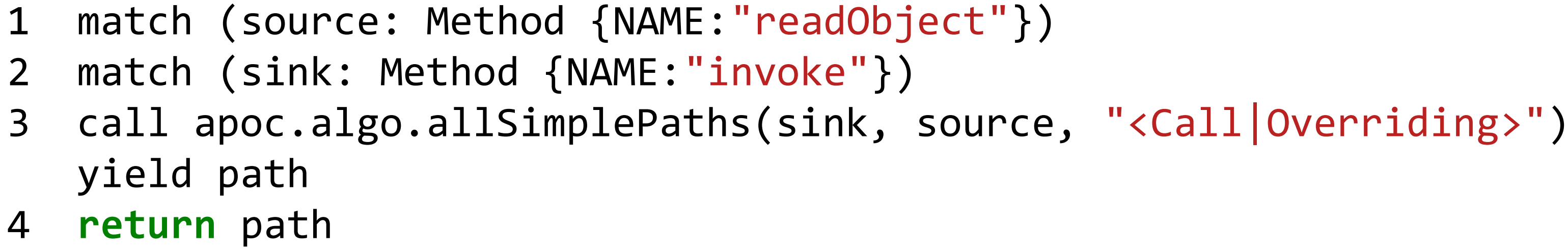}
\caption{A simple query script example.}
\label{search}
\end{figure}

%\begin{figure}
%\begin{minted}[numbersep=5pt,escapeinside=||,linenos,frame=leftline,breaklines]{java}

%match (source: Method {NAME:"readObject"})
%match (sink: Method {NAME:"invoke"})
%call apoc.algo.allSimplePaths(sink, source, "<CALL|ALIAS") %yield path 
%return path
%\end{minted}
%\caption{A simple query script example}
%\label{search}
%\end{figure}

\noindent\textbf{Example.} Figure \ref{search} presents an example of our query scripts for suspicious gadget chain identification. We first select the magic methods and security-sensitive call sites we are interested in (line 1-2). For example, we adopt \code{readObject()} and \code{invoke()} as the source and sink respectively, and use the built-in method \code{allSimplePaths()}\footnote{https://neo4j.com/developer/neo4j-apoc/} (line 3) to search for all suspicious gadget chains (i.e., reachable paths from the target source to the sink). A set of paths, which satisfy the following three conditions: 1) its start node is a source method; 2) its end node is a sink method; and 3) the type of connected edges is \code{CALL} or \code{Overriding}, is returned as our retrieval results (line 4), i.e., suspicious gadget chains.

\subsection{Chain Verification}\label{Chains Verification}
\begin{figure}
  \centering
  \includegraphics[width=.9\linewidth]{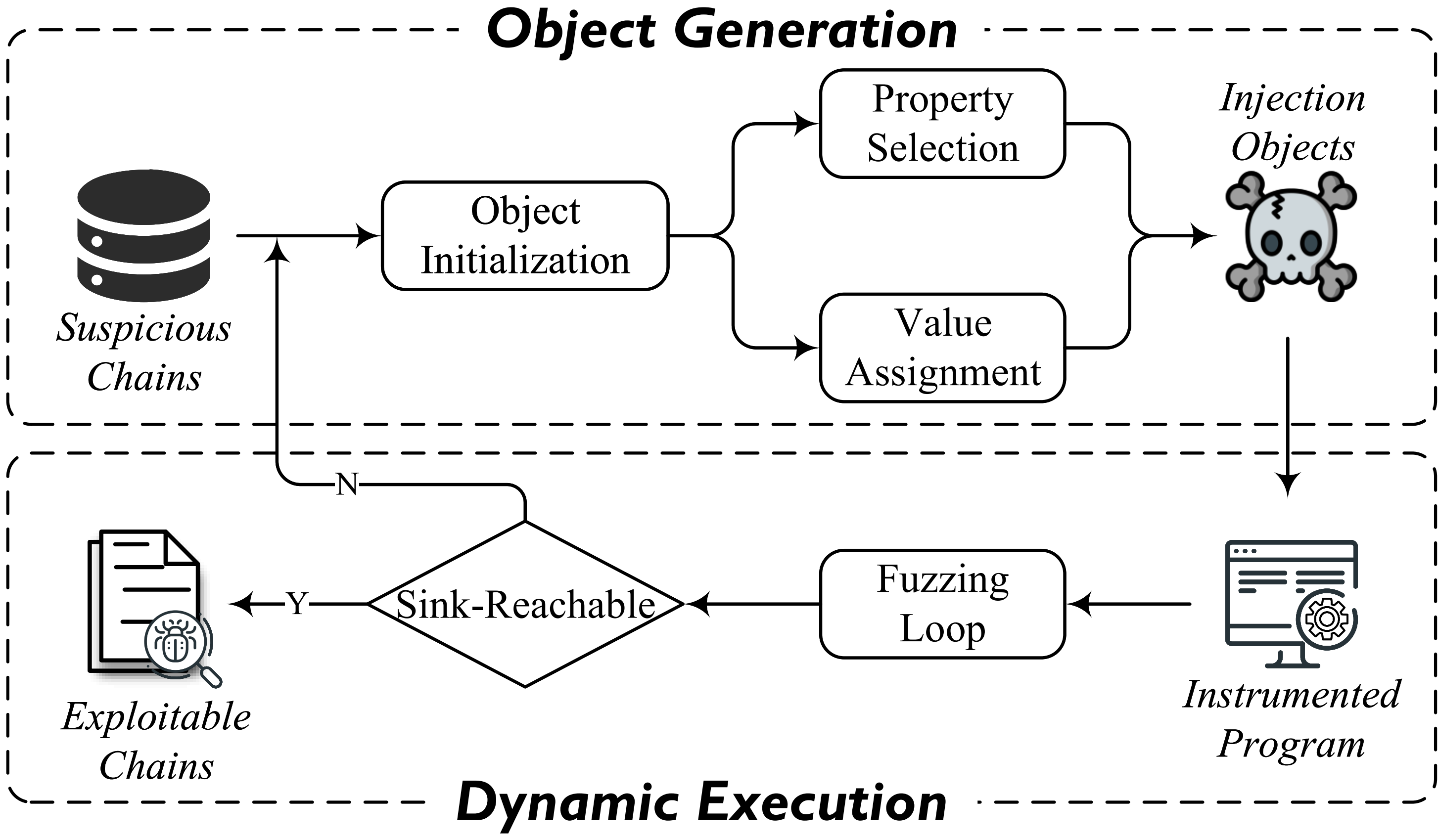}
\caption{Overview of gadget chain verification.}
\label{Fuzzing Loop}
\end{figure}

Given a set of suspicious gadget chains, \emph{GCMiner} leverages an overriding-guided object generation approach to produce valid injection objects and performs coverage-guided fuzzing for verification. Figure \ref{Fuzzing Loop} shows the overview of our chain verification, which contains two modules: 1) \emph{Object Generation}; and 2) \emph{Dynamic Execution}. For object generation, \emph{GCMiner} first selects a gadget chain from candidates and instantiates the class to which the given chain's first gadget (i.e., deserialization entry point, or magic method) belongs to construct an initial injection object. Then, to enable the generated injection object to follow the execution flow of the target gadget chain, \emph{GCMiner} modifies the property values of the injection object by dynamic binding. For dynamic execution, \emph{GCMiner} feeds these generated injection objects into the target program for fuzzing. Once an injection object reaches the target security-sensitive call site, this gadget chain under testing will be confirmed as exploitable. The procedure of chain verification will not terminate until all statically identified candidate chains have been checked.

\noindent\textbf{Object Generation.}
Generating an injection object for a given gadget chain requires dynamically modifying its property values to enable the injection object to follow the execution flow of the identified gadget chain to reach security-sensitive call site. To achieve this, we leverage overriding relations between methods to guide injection object generation.

Specifically, \emph{GCMiner} first generates an initial injection object by instantiating the class to which the entry point (i.e., magic method) of the target chain belongs. Then, according to the relation information provided by DA-CG, \emph{GCMiner} modifies the property values of the initial injection object through dynamic binding to ensure that the injection object can reach the dangerous sink. Such a dynamic property value assignment requires 1) selecting a property, which receives a serializable class object as its value, from the property list of the injection object (i.e., \emph{property selection}); and 2) assigning an instantiated  object as the value to the corresponding property (i.e., \emph{value assignment}). 

For property selection, we dynamically obtain each property type of the injection object by Java reflection \cite{reflection1,ref1,ref2}, which allows a software system to inspect and change the behavior of its classes, interfaces, methods, and fields at runtime. When a property type is a class object, this property is listed as a candidate for value assignment. These candidate properties may be exploited by attackers to implement insecure deserialization paths. For value assignment, we adopt the overriding edge provided by our constructed DA-CG as guidance to select an overridden method that can be invoked by dynamic binding to enable the injection object to reach the dangerous sink of the given gadget chain. \emph{GCMiner} instantiates the class to which the overridden method belongs, and analyzes whether the class object can be assigned as the property value to a candidate property of the injection object. If the class to which the overridden method belongs is a subclass of the property type of the injection object, \emph{GCMiner} will assign the instance containing the overridden method to the corresponding property of the injection object.

Considering that the properties of a given gadget chain may need multiple modification to facilitate the execution of the gadget chain, we repeat the above process (i.e., property selection and value assignment) for dynamic property value assignment to enable the injection object to reach the security-sensitive call site. Once the property values of the injection object are configured, \emph{GCMiner} will feed it into the target application for fuzzing.

\noindent\textbf{Dynamic Execution.}
As discussed before, an exploitable gadget chain should enable the malicious deserialized object to reach (in terms of control flow) and affect (in terms of data flow) the security-sensitive sink, while our overriding-guided object generation strategy just assure the reachability of these statically reported gadget chains (i.e., attackers can dynamically invoke these gadgets to inject malicious objects). Hence, to verify whether the dangerous sink of the given gadget chain can be affected by the generated injection object, \emph{GCMiner} adopts a generation-based coverage-guided Java fuzzing framework JQF \cite{JQF}, which provides an extensible interface for users to easily integrate their own input generation mechanism for testing. 

Specifically, \emph{GCMiner} first executes the injection object on the instrumented program and collects the code coverage as feedback to guide the fuzzing procedure. The injection object which covers more branches in the gadgets is more likely to reach the target sink. Unlike traditional coverage-guided fuzzing which blindly increases the code coverage to accidentally trigger potential vulnerabilities, \emph{GCMiner} only instruments classes to which gadgets belong on the application's classpath, which makes the fuzzer pay more attention to those seeds (i.e., generated or mutated injection objects) that trigger more code snippets within gadgets. These seeds are more likely to reach dangerous sinks. For mutation, \emph{GCMiner} leverages the JQF-Zest algorithm \cite{Zest} to produce new inputs that get deeper into the target gadget chain by mutating the interesting seeds at the bit-level. These bit-level mutations correspond to property-level mutations on structured injection objects \cite{DBLP:journals/ese/ZhouBWSZLZC22}. For \emph{primitive} data types (e.g., \code{boolean}, \code{int}), the fuzzer uses multiple pseudo-random methods built in JQF to convert untyped bit parameters into random typed values. For the \emph{reference} data types, the fuzzer tailors targeted templates for specific types. When the property type is \code{class}, the fuzzer will randomly select a class from the candidate classes (i.e., sub-classes) of this property via the method \code{random.choose()}. For an \code{array} property, the fuzzer uses the method \code{random.nextInt()} to randomly set up the array size and assigns random values based on the type of elements (i.e., instances that inherit the class type of the array) to the array.

Once a generated injection object reaches the target security-sensitive call site, the suspicious chain under testing will be confirmed as a gadget chain. The procedure of chain verification will not terminate until all the statically reported gadget chains have been checked.

\section{Experiment}\label{Experiments}
\subsection{Research Questions}

\noindent\textbf{RQ3: Effectiveness of \emph{GCMiner}. }How effective is \emph{GCMiner} in mining Java deserialization gadget chains?
\
\newline
\indent
By investigating this RQ, we aim to answer how well does \emph{GCMiner} perform in comparison with the state-of-the-art automatic Java deserialization gadget chain mining techniques.

\noindent\textbf{RQ4: Ablation study.}

\hspace{-1em}\textbf{- RQ4a: Impact of additional sources and sinks. }Whether these newly added sources and sinks contribute to mining more exploitable gadget chains?
\
\newline
\indent
Our approach includes some new magic methods and security-sensitive call sites, which aims to search for more suspicious gadget chains. This question aims to show whether these additional sources and sinks can help find more exploitable gadget chains?

\hspace{-1em}\textbf{- RQ4b: Impact of introducing method overriding. }Does the introduction of overriding relations contribute to identifying more exploitable gadgets?
\
\newline
\indent
One of our key insights is to introduce additional overriding relations to capture dynamic program features abused by attackers for gadget chain construction. By investigating this RQ, we aim to show whether the introduction of overriding relations helps  identify more exploitable gadget chains.

\hspace{-1em}\textbf{- RQ4c: Impact of overriding-guided object generation. }Can our overriding-guided object generation approach produce valid injection objects for fuzzing?
\
\newline
\indent
Another key insight of our approach is that generating exploitable injection objects through binding exploitable gadget dynamically to enable them to reach dangerous sinks. By answering this RQ, we aim to show whether leveraging overriding relations to guide dynamic binding can generate valid injection objects for gadget chain verification.

\subsection{Experiment Setup}
\subsubsection{Benchmark} 
To evaluate the effectiveness of \emph{GCMiner}, we adopted our manually constructed Java deserialization vulnerability benchmark in Table \ref{data}, which consists of a widely-used gadget chain collection ysoserial \cite{ysoserial} repository and 18 famous Java applications with 86 known gadget chains.

\subsubsection{Baseline methods}
We compared \emph{GCMiner} with a well-known open-source tool, \emph{Gadget Inspector} \cite{Inspector}, and a previous study, \emph{Serhybrid} \cite{Serhybrid}. %\emph{Gadget Inspector} leverages static taint analysis and simple symbolic execution to mine the propagation paths of parameters within/between methods of a target application, and then performs a Breadth-first search (BFS) to search for attacker-controllable gadget chains. \emph{Serhybrid} first combines points-to and heap access path analysis to search for tainted data which can flow from a deserialized object to the security-sensitive call site. Then,  \emph{Serhybrid} adopts fuzzing to generate actual input objects for automatic verification of the identified gadget chains.

\subsubsection{Implementation}
We used \emph{Tabby} \cite{Tabby}, a Java code analysis tool based on \emph{Soot} \cite{Soot}, to extract both call and overriding relations between methods for DA-CG construction. For chain identification, we used a popular graph database Neo4j \cite{Neo4j} to perform our customized query scripts. To verify statically identified gadget chains, we implemented our overriding-guided object generation strategy based on JQF \cite{JQF}, a coverage-guided Java fuzzing framework. JQF was selected for its extensibility in implementing structured seed generation templates. All experiments were conducted on a Linux workstation with an Intel(R) Core(TM) i9-12900k @3.90GHz and 128 GB of RAM, running Ubuntu 18.04.4 LTS with JDK 1.8.0\_152. 

\subsubsection{Experimental configurations}
For RQ3, we ran \emph{GCMiner} and baselines on vulnerability-specific versions of applications in our benchmark. Unfortunately, despite our best efforts, \emph{Serhybrid} was not reproducible. We made unsuccessful attempts to contact the authors for suggestions. We could only compare \emph{GCMiner} with \emph{Serhybrid} on the results of several applications reported in the original paper. To ensure the fairness of the comparison, we conducted the experiments under the same conditions and evaluated the performance with the same metrics. We repeated each experiment 10 times and reported their average performance \cite{DBLP:conf/ccs/KleesRCW018}. According to the assessment of our employed security experts (each of them had two to five years of vulnerability mining-related experience gained in industry), we empirically set the threshold for the length of each chain to 15 gadgets to avoid the path explosion problem during graph traversal. For each statically identified gadget chain, we limit the fuzzing campaign of \emph{GCMiner} to 120 seconds. For RQ4a, we only used statically constructed CGs for gadget chain identification to evaluate the contribution of overriding relations to capturing exploitable gadgets. In RQ4b, to investigate the contribution of our overriding-guided object generation, we randomly built injection objects based on DA-CG for comparison.

\subsection{Evaluation Metrics}\label{metric}
To evaluate our approach, we used the following  metrics.

\textbf{\emph{Known Gadget Chains} (KGC) }is the number of the publicly known gadget chains in a target application.

\textbf{\emph{Reported Gadget Chains} (Rep) }computes the total number of gadget chains statically reported by each approach.

\textbf{\emph{True Positives} (TP) }is the number of truly exploitable gadget chains reported by each approach. In our experimental evaluation, TP counts how many known gadget chains in the benchmark are mined.

\textbf{\emph{Precision} (P) }is the fraction of truly exploitable gadget chains among the reported ones. It is calculated as: $P = \frac{TP}{Rep}$.

\textbf{\emph{Recall} (R) }is the fraction of known gadget chains that are identified by each approach. It is calculated as: $R = \frac{TP}{KGC}$.

%\subsection{Experimental Results}\label{Results}
\subsection{Effectiveness of GCMiner (RQ3)}

\begin{table}[t]
  \centering
  \fontsize{5.5}{10}\selectfont
  \caption{Comparison results between \emph{GCMiner} and \emph{Gadget Inspector}.}
  \begin{threeparttable}
    \begin{tabular}{|l|c|c|c|c|c|c|c|}
    \hline
    \multirow{2}{*}{\textbf{Application}} & \multirow{2}{*}{\textbf{\#KGC}} &  \multicolumn{3}{c|}{\textbf{\emph{GCMiner}}}&\multicolumn{3}{c|}{\textbf{\emph{Gadget Inspector}}}\cr\cline{3-8}
    ~ & ~ &\textbf{\#TP/\#Rep}&\textbf{P\tnote{*}}&\textbf{R}&\textbf{\#TP/\#Rep}&\textbf{P}&\textbf{R}\cr  
    \hline
    \hline
    ysoserial &  34  & 21 / 29  & 1  & 0.618 & 3 / 116 & 0.026 &0.088\cr
    \hline
    \hline
    JBoss RESTEasy  & 1 & 1 / 3  & 1 & 1 & 0 / 2 & 0&0\cr
    Apache Camel  & 2 & 2 / 2  &  1 & 1 &  0/ 2  & 0&0\cr
    Apache Brooklyn & 1 & 1 / 1 &1 & 1 & 0 / 2 & 0 & 0\cr
    Apache XBean  &  1  & 0 / 2  & 1 & 0 & 0 / 2& 0&0 \cr
    Shiro &  3& 1 / 2 & 1 & 0.333 & 0 / 2&0&0\cr
    Pippo  &2 & 2 / 5 & 1 & 1 & 0 / 2 &0& 0\cr
    Adobe Coldfusion & 2& 2 / 3  & 1& 1&  1/ 2& 0.500 &0.500\cr
    VMWare VCenter  &  1 & 1 / 1 &1& 1 &0 / 2 &  0 &0\cr
    Red5 &   1 & 1 / 2 & 1& 1 & 0/2 &   0 &0\cr
    Hessian &5 & 4 / 7  & 1& 0.800 &   0 / 2 &0 &0\cr
    XStream &14 & 12 / 19  & 1 & 0.857 &   1 / 2& 0.500 &0.071\cr
    Commons Collections &3  & 3 / 7 & 1& 1 & 0 / 12 &0& 0\cr
    Dubbo & 2 & 1 / 2 & 1 & 0.500& 0 / 3& 0&0\cr
    WebLogic &5& 4 / 11 & 1& 0.800 &  0 / 6 &0 &0\cr
    Emissary& 3 & 2 / 4  & 1 & 0.667& 0 / 3 & 0&0\cr
    Jenkins& 2& 1 / 9  & 1 & 0.500 &   0 / 2 & 0&0\cr
    Apache OFBiz & 3 & 1 / 4 & 1 &0.333 & 0 / 2  & 0& 0\cr
    Spring  & 1  & 1 / 5 & 1& 1 &  0 / 6 &0 &0\cr
    \hline
    \textbf{Total} &  \textbf{86} & \textbf{61 / 118} &\textbf{1} & \textbf{0.709}& \textbf{5 / 172} & \textbf{0.029}& \textbf{0.058}\cr\cline{1-2}
    \hline
    \end{tabular}
    \begin{tablenotes}  
    \footnotesize         
    \item[*] Since \emph{GCMiner} adopted fuzzing to verify exploitable gadget chains, we used dynamically confirmed gadget chains as \emph{Rep} to compute the precision.
    \end{tablenotes}
    \end{threeparttable}
    \label{RQ3}
\end{table}

Table \ref{RQ3} shows the overall results of \emph{GCMiner}. In total, \emph{GCMiner} identifies 61 out of 86 known gadget chains with a recall of $61/86 = 70.9\%$ without false positives. The reasons for these false negatives mainly come from two aspects. In the static identification stage, due to the limited support for certain dynamic features of Java language such as \emph{reflective calls} \cite{reflection1} and \emph{dynamic proxy} \cite{DBLP:conf/issta/FourtounisKS18}, certain exploitable gadgets on the classpath of the application are not captured by our DA-CG. For example, in \code{Groovy1} \cite{Groovy1}, the attacker could exploit the class \code{ConvertedClosure}, whose constructor receives a proxy \code{MethodClosure} as its parameters, to pass tainted arguments to the gadget \code{MethodClosure.call()} to execute the malicious commands. Due to the unawareness of which classes can be proxied, gadget chains involving dynamic proxy during their construction are difficult to be identified by \emph{GCMiner}, resulting in false negatives. In the dynamic verification stage, due to certain specific constraints, some statically identified gadget chains cannot be dynamically validated by \emph{GCMiner}. For example, \emph{GCMiner} fails to construct sink-reachable injection objects for \code{AspectJWeaver} \cite{AspectJWeaver} in \emph{ysoserial} because its sink method \code{writeToPath()} receives a file as an input, which is not supported by our injection object generation strategy.

\begin{figure}
  \centering
  \includegraphics[width=\linewidth]{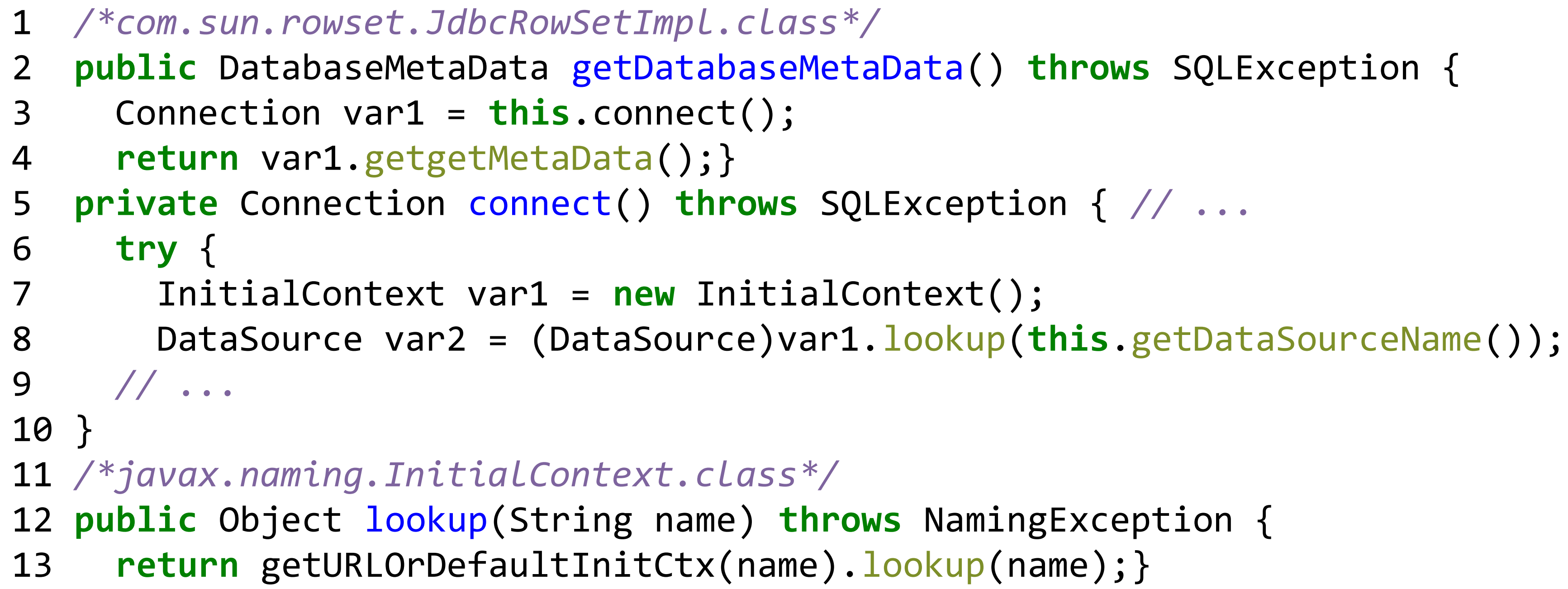}
\caption{A gadget chain identified by \emph{GCMiner} but missed by \emph{Gadget Inspector}.}
\label{RQ3Example}
\end{figure}

\underline{\emph{GCMiner} vs. \emph{Gadget Inspector}.}
As shown in Table \ref{RQ3}, \emph{GCMiner} identifies a total of 118 suspicious gadget chains, of which 61 are known gadget chains. By contrast, \emph{Gadget Inspector} can only identify five exploitable gadget chains with a recall of $5/86=5.8\%$, and a precision of $5/172=2.9\%$. In addition, all gadget chains identified by \emph{Gadget Inspector} are covered by \emph{GCMiner}. Such a significant performance gap may result from two aspects. On the one hand, constrained by a limited number of exploitable magic methods and security-sensitive call sites, a large number of suspicious gadget chains on the classpath of the application are missed by \emph{Gadget Inspector}. Figure \ref{RQ3Example} presents a typical example which can be identified by \emph{GCMiner} but missed by \emph{Gadget Inspector}. It is a widely exploited gadget chain which can be triggered to perform JNDIi attack. During the process of object deserialization, the method \code{getDatabaseMetaData()} (line 2) in class \code{JdbcRowSetImpl} will invoke the method \code{connect()} (line 3) by default. To get the context of the object transferred by users, the method \code{connect()} will invoke the method \code{lookup()} (line 8), which is a security-sensitive call site that can be exploited by remote attackers to inject malicious code. However, since the method \code{lookup()} is not considered by \emph{Gadget Inspector}, this gadget chains is missed. On the other hand, due to the limited precision of static analysis, \emph{Gadget Inspector} cannot guarantee that identified gadget chains are truly exploitable, resulting in many \emph{false positives}. Owing to our overriding-guided object generation approach which produces valid injection objects for verification, the gadget chains which cannot be exploited  will be filtered out.

\underline{\emph{GCMiner} vs. \emph{Serhybrid}.}
We also compared \emph{GCMiner} with \emph{Serhybrid} on the selected applications from \emph{ysoserial} repository. Table \ref{GVS} presents the comparative results of \emph{GCMiner} and \emph{Serhybrid}. Column ``\#Object" presents the number of injection objects generated by each approach. Column ``\#Exploit" is the number of injection objects that can trigger known gadget chains.

\begin{table}[t]
  \centering
  \fontsize{6.3}{8}\selectfont
  \caption{Comparison results between \emph{GCMiner} and \emph{Serhybrid}.}
    \begin{tabular}{|l|c|c|c|c|c|}
    \hline
    \multirow{2}{*}{\textbf{Application}} & \multirow{2}{*}{\textbf{\#KGC}} &  \multicolumn{2}{c|}{\textbf{\emph{GCMiner}}}&\multicolumn{2}{c|}{\textbf{\emph{Serhybrid}}} \cr\cline{3-6}
    ~ & ~ &\textbf{\#Object}&\textbf{\#Exploit}&\textbf{\#Object}&\textbf{\#Exploit}\cr  
    \hline
    \hline
    bsh-2.0b5  & 1 & 1 & 0 & 0 & 0 \cr
    clojure-1.8.0  & 1  & 2 & 1 & N/A  & 0 \cr
    commons-beanutils-1.9.2 & 1 & 2 & 1 & 0 & 0\cr
    commons-collections-3.1  & 5 &  12 & 3 & 1  &  1\cr
    commons-collections4-4.0 & 2 & 4  & 2 & 1 &  1\cr
    groovy-2.3.9  & 1  & 2 & 0 & 0 & 0\cr
    hibernate & 2 & 3 & 2 &0 &0\cr
    jython-standalone-2.5.2  & 1  &  1 & 0 & N/A& 0\cr
    rome-1.0 & 1   &  2  & 1  & 0 &0\cr
    \hline
    \textbf{Total} &  \textbf{15} & \textbf{29} &\textbf{10} & \textbf{2} & \textbf{2}\cr
    \hline
    \end{tabular}
    \label{GVS}
\end{table}

The results show that, from the nine applications which contain 15 known gadget chains, \emph{GCMiner} successfully generates 29 injection objects, 10 of which are valid injection objects that can be leveraged by attackers to perform deserialization attack. By contrast, \emph{Serhybrid} generates two valid injection objects. Both two gadget chains reported by \emph{Serhybrid} are covered by \emph{GCMiner}. Although \emph{Serhybrid} achieves higher precision due to its points-to analysis, many gadget chains are missed. The reason may be that the heavy use of Java dynamic program features makes \emph{Serhybrid} hard to compute reachable paths to dangerous sinks. As a result, only a small number of objects can be successfully generated  for verification. By contrast, owing to our constructed DA-CG, \emph{GCMiner} can better model complex dynamic behaviors of programs and contribute to capturing more exploitable gadgets. Besides, from the results, we can observe that \emph{Serhybrid} is hard to generate valid injection objects (timeout even occurred when analyzing \code{Clojure} \cite{Clojure} and \code{Jython} \cite{Jython}) for verification. Such situation occurs  due to the strict object generation strategy of \emph{Serhybrid} that relies on the analysis results of heap access paths. Therefore, once the precise execution paths which perform malicious objects to the target sink cannot be identified by static analysis, \emph{Serhybrid} is difficult to produce valid injection objects.

\begin{tcolorbox}[left=1pt,right=1pt,top=1pt,bottom=1pt,boxrule=0.7pt,enhanced,drop fuzzy shadow]
\textbf{Answer to RQ3:}\textit{ GCMiner significantly outperforms the state-of-the-art Java deserialization gadget chain mining tools, identifying 56 unique gadget chains that cannot be identified by baselines.}
\end{tcolorbox}

\subsection{Ablation study (RQ4)}\label{Ablation}

\begin{table}[t]
  \centering
  \fontsize{5.3}{8}\selectfont
  \caption{Impact of additional sources and sinks on gadget chain mining.}
    \begin{tabular}{|l|c|c|c|c|c|c|c|}
    \hline
    \multirow{2}{*}{\textbf{Application}} & \multirow{2}{*}{\textbf{\#KGC}} &  \multicolumn{2}{c|}{\textbf{\emph{GCMiner}}}&\multicolumn{2}{c|}{\textbf{\emph{GCMiner$_{Var}$}}}&\multicolumn{2}{c|}{\textbf{\emph{Gadget Inspector$_{Var}$}}}\cr\cline{3-8}
    ~ & ~ &\textbf{\#Rep}&\textbf{\#TP}&\textbf{\#Rep}&\textbf{\#TP}&\textbf{\#Rep}&\textbf{\#TP}\cr  
    \hline
    \hline
    ysoserial &  34  & 29  & 21   & 24  & 15 & 637 & 4\cr
    \hline
    \hline
    JBoss RESTEasy  & 1 & 3  & 1 &  2  & 1  & 14 & 0\cr
    Apache Camel  & 2 & 2  &  2  &  2 &  2 &  14 &0\cr
    Apache Brooklyn & 1 & 1 &1 & 1 & 1 & 16&0\cr
    Apache XBean  &  1  & 2  & 0 & 1  &  0 & 14 &0\cr
    Shiro &  3& 2 & 1 & 1 & 0 & 14  &0\cr
    Pippo  &2 & 5 & 2& 3 & 1 &14&0\cr
    Adobe Coldfusion & 2& 3  & 2& 3 & 2  & 14 & 1\cr
    VMWare VCenter  &  1 & 1 &1  & 1 & 1  &12& 0\cr
    Red5 &   1 & 2 & 1& 1 &  1 & 14&0 \cr
    Hessian &5 & 7  & 4&  5  & 3 &14 & 0\cr
    XStream &14 & 19  & 12 & 15  & 10 & 14 & 2\cr
    Commons Collections &3  & 7 & 3& 7  & 3 &  69 &0\cr
    Dubbo & 2 & 2 & 1  & 2 & 1 & 16& 0\cr
    WebLogic &5& 11 & 4 &  8 & 3 & 21 & 0\cr
    Emissary& 3 & 4  & 2  & 3 & 2 & 11 &0\cr
    Jenkins& 2& 9  & 1 &  6 & 1 &14 &0\cr
    Apache OFBiz & 3 & 4 & 1 & 2 & 1 & 14 &0\cr
    Spring  & 1  & 5 & 1 & 4 &  1 & 46 &0 \cr
    \hline
    \textbf{Total} &  \textbf{86} & \textbf{118} &\textbf{61} & \textbf{91} & \textbf{49}& \textbf{982}& \textbf{7}\cr\cline{1-2}
    \hline
    \end{tabular}
    \label{RQ4a}
\end{table}

\subsubsection{\textbf{RQ4a: Impact of additional sources and sinks}}
Table \ref{RQ4a} summarizes the number of known gadget chains discovered by \emph{GCMiner} and other variants. Overall, with these additional sources and sinks newly collected in our previous empirical study, \emph{GCMiner} identifies 27 more probably exploitable gadget chains, 12 of which are the  known gadget chains. Besides, we also notice that the improvement of \emph{Gadget Inspector$_{Var}$} is not significant. Despite of additional sources and sinks, \emph{Gadget Inspector$_{Var}$} only mines two more exploitable gadget chains but reports 808 more false positives. The reason may be that the introduction of new exploitable sources and gadgets can improve the search scope about suspicious gadget chains to a certain extent. Meanwhile, it also amplifies the deficiencies in imprecision of static analysis in gadget chain mining. By contrast, our reflection-guided exploit generation approach can effectively filter out invalid gadget chains.

\begin{tcolorbox}[left=1pt,right=1pt,top=1pt,bottom=1pt,boxrule=0.7pt,enhanced,drop fuzzy shadow]
\textbf{Answer to RQ4a:}\textit{ Additional exploitable magic methods and security-sensitive call sites are useful to identify more potential gadget chains.}
\end{tcolorbox}

\subsubsection{\textbf{RQ4b: Impact of introducing method overriding}}
Table \ref{RQ4b} shows the number of gadget chains identified by \emph{GCMiner} in different configurations. Columns ``With Overriding" and ``W/O Overriding" represent \emph{GCMiner} enabling/disabling method overriding in DA-CG, respectively. The results demonstrate that the introduction of method overriding positively contributes to improving the effectiveness of \emph{GCMiner}. In particular, by taking method overriding into consideration, \emph{GCMiner} can identify 118 suspicious gadget chains, of which 61 are true positives. By contrast, \emph{GCMiner} based on statically constructed CGs can only identify 3 out 9 exploitable gadget chains. Hence, overriding relations can effectively capture implicit method invocations and is helpful for mining exploitable gadget chains.

\begin{table}[t]
  \centering
  \fontsize{7.95}{10}\selectfont
  \caption{Impact of introducing overriding relations on gadget chain identification.}
    \begin{tabular}{|l|c|c|c|c|c|}
    \hline
    \multirow{2}{*}{\textbf{Application}} & \multirow{2}{*}{\textbf{\#KGC}} &  \multicolumn{2}{c|}{\textbf{\emph{With Overriding}}}&\multicolumn{2}{c|}{\textbf{\emph{W/O Overriding}}} \cr\cline{3-6}
    ~ & ~ &\textbf{\#Rep}&\textbf{\#TP}&\textbf{\#Rep}&\textbf{\#TP}\cr  
    \hline
    \hline
    ysoserial &  34  & 29 & 21  & 6  &  2 \cr
    \hline
    \hline
    JBoss RESTEasy  & 1 & 3  & 1 & 0 & 0\cr
    Apache Camel  & 2 & 2  &  2 & 1  & 0\cr
    Apache Brooklyn & 1 & 1 & 1 & 0 & 0 \cr
    Apache XBean  &  1  & 2  & 0 & 0 & 0 \cr
    Shiro &  3& 2 &  1 & 0 &0\cr
    Pippo  &2 & 5 & 2 & 1 & 0\cr
    Adobe Coldfusion & 2& 3  &2 & 0 &  0\cr
    VMWare VCenter  &  1 & 1 &1& 0& 0  \cr
    Red5 &   1 & 2 & 1& 0 &  0  \cr
    Hessian &5 & 7  &4 & 0 & 0\cr
    XStream &14 &  19 & 12 & 3  & 0\cr
    Commons Collections &3  & 7 & 3 & 2 & 1\cr
    Dubbo & 2 & 2 & 1 & 0 &0 \cr
    WebLogic &5& 11  & 4 & 1 & 0\cr
    Emissary& 3 & 4  &  2 & 0 & 0\cr
    Jenkins& 2& 9  & 1 &  1 & 0\cr
    Apache OFBiz & 3 & 4 & 1 & 0 &  0\cr
    Spring  & 1  & 5 & 1 &  0 & 0\cr
    \hline
    \textbf{Total} &  \textbf{86} & \textbf{118} &\textbf{61} & \textbf{9} & \textbf{3}\cr
    \hline
    \end{tabular}
    \label{RQ4b}
\end{table}

\begin{tcolorbox}[left=1pt,right=1pt,top=1pt,bottom=1pt,boxrule=0.7pt,enhanced,drop fuzzy shadow]
\textbf{Answer to RQ4b:}\textit{ The introduction of overriding relations significantly enhances the capability of existing static analysis in capturing potential exploitable gadgets, enabling our approach to identify more exploitable gadget chains.}

\end{tcolorbox}

\subsubsection{\textbf{RQ4c: Impact of overriding-guided object generation}}

\begin{table}[t]
  \centering
  \fontsize{6.8}{8}\selectfont
  \caption{Impact of overriding-guided object generation on gadget chain verification.}
    \begin{tabular}{|l|c|c|c|c|c|}
    \hline
    \multirow{2}{*}{\textbf{Application}} & \multirow{2}{*}{\textbf{\#KGC}} &  \multicolumn{2}{c|}{\textbf{\emph{GCMiner}}}&\multicolumn{2}{c|}{\textbf{\emph{GCMiner$_{NG}$}}} \cr\cline{3-6}
    ~ & ~ &\textbf{\#Object}&\textbf{\#Exploit}&\textbf{\#Object}&\textbf{\#Exploit}\cr  
    \hline
    \hline
    ysoserial &  34  & 86 & 21  & 5  &  0 \cr
    \hline
    \hline
    JBoss RESTEasy  & 1 & 3  & 1 & 0 & 0\cr
    Apache Camel  & 2 & 7  &   2 & 0  & 0\cr
    Apache Brooklyn & 1 & 3 & 1 & 0 & 0 \cr
    Apache XBean  &  1  & 2  & 0 & 0 & 0 \cr
    Shiro &  3& 6 &  1 & 0 &0\cr
    Pippo  &2 & 5 & 2 & 0 & 0\cr
    Adobe Coldfusion & 2& 7  &2 & 0 &  0\cr
    VMWare VCenter  &  1 & 3 &1& 0& 0  \cr
    Red5 &   1 & 2 & 1& 0 &  0  \cr
    Hessian &5 & 11  &4 & 0 & 0\cr
    XStream &14 &  48 & 12 & 1  & 0\cr
    Commons Collections &3  & 8 & 3 & 1 & 0\cr
    Dubbo & 2 & 4 & 1 & 0 &0 \cr
    WebLogic &5& 13  & 4 & 0 & 0\cr
    Emissary& 3 & 9  &  2 & 0 & 0\cr
    Jenkins& 2& 3  & 1 &  0  & 0\cr
    Apache OFBiz & 3 & 5 & 1 & 0 &  0\cr
    Spring  & 1  & 4 & 1 &  0 & 0\cr
    \hline
    \textbf{Total} &  \textbf{86} & \textbf{229} &\textbf{61} & \textbf{7} & \textbf{0}\cr
    \hline
    \end{tabular}
    \label{RQ4c}
\end{table}

Table \ref{RQ4c} presents the  results of \emph{GCMiner} and its corresponding variant (\emph{GCMiner$_{NG}$}) which generates injection objects with no guidance during fuzzing. The results show that our overriding-guided object generation can effectively generate injection objects, 61 of which are valid exploits. By contrast, \emph{GCMiner$_{NG}$} can only generate seven objects, none of which can reach the dangerous sink. This performance gap may be due to  that constrained by the highly structured characteristic (i.e., the property values of an exploitable object need to be modified many times to reach the dangerous sink) of the injection object, existing object generation techniques can hardly produce valid injection objects for verification.

\begin{tcolorbox}[left=1pt,right=1pt,top=1pt,bottom=1pt,boxrule=0.7pt,enhanced,drop fuzzy shadow]
\textbf{Answer to RQ4c:}\textit{ With our overriding-guided object generation, GCMiner can effectively generate sink-reachable injection objects for fuzzing.}
\end{tcolorbox}

\section{Threats to Validity}\label{Threats}
\subsection{External validity}
A main external threat comes from the generalization of our empirical results. Similar to existing works \cite{Inspector,Serhybrid,FUGIO,DBLP:conf/ccs/DahseKH14}, the effectiveness (recall) of our static gadget chain identification also relies heavily on the prior expert knowledge of available sources and sinks. Considering that there are a few orthogonal tools/approaches \cite{Scanner,objectMap} have been proposed to automatically identify untrusted deserialization entry points, and our knowledge base is configurable, i.e., newly disclosed sources and sinks can be dynamically added, the capability to detect unknown Java deserialization vulnerabilities in the wild can be improved. Furthermore, since we only focus on Java deserialization vulnerabilities, all findings and evaluation results may not be applicable to other programming languages (e.g., PHP \cite{FUGIO} and .NET \cite{shcherbakov2021serialdetector}) which also suffer from the risk of deserialization vulnerabilities. We leave it as a future work to extend our approach to other languages.

\subsection{Internal validity}
Internal validity in our experiment comes from two aspects. On the one hand, since our approach aims to identify exploitable gadget chains instead of detecting Java deserialization vulnerabilities, gadget chains which cannot be exploited in practice may be wrongly reported. To avoid the bias in our conclusions, we tried our best to manually reproduce each vulnerability in our benchmark to make sure the practical exploitability of gadget chains. On the other hand, due to the non-reproducibility of \emph{Serhybrid}, we directly compared our approach with \emph{Serhybrid} on its reported applications and results under the same experimental situations. 

\section{Related Work}\label{related work}

\textbf{Java Deserialization Vulnerability Detection.}
Many studies have been proposed for analyzing, defensing, and detecting Java deserialization vulnerabilities \cite{DBLP:conf/uss/AzadLN19,DBLP:conf/icse/CaoSBWLT22,DBLP:journals/infsof/CaoSBWL21,DBLP:journals/infsof/ZhouSXLC19,DBLP:journals/iet-sen/SubhanWBSR22}. Muñoz et al. \cite{Munoz} conducted a comprehensive analysis on JSON deserialization libraries and presented several mitigation measures as takeaways. Carettoni \cite{Carettoni} presented a configurable Java deserialization library, which supports multiple optional settings such as blacklist and whitelist, to secure application from untrusted input. Koutroumpouchos et al. \cite{objectMap} proposed an extendable tool \emph{ObjectMap}, which generates a series of requests to validate whether the payload can be directly passed to the target application. Cristalli et al. \cite{DBLP:conf/raid/CristalliVBL18} proposed a dynamic approach, which collects the behavior information of benign deserialization process and constructs the precise execution path to prevent untrusted data input. Sayar et al. \cite{10.1145/3554732} conducted a large-scale empirical study on publicly known Java deserialization RCE exploits and investigated how deserialization vulnerabilities manifest in real code bases and libraries.

In order to automatically mine suspicious gadget chains, Haken \cite{Inspector} presented \emph{Gadget Inspector}, which leverages static taint analysis and simple symbolic execution to mine the propagation paths of parameters within/between methods of a target application, and then performs a Breadth-first search (BFS) to search for exploitable gadget chains. Rasheed et al. \cite{Serhybrid} proposed \emph{Serhybrid}, a hybrid analysis-based approach which constructs a heap abstraction to produce actual input objects to automatically verify exploitable gadget chains. In this paper, \emph{GCMiner} constructs DA-CG to identify more exploitable gadget chains, and leverages an overriding-guided object generation approach to produce valid injection objects.

\textbf{Automatic Exploit Generation}. 
Automatic Exploit Generation (AEG) is proposed to automatically construct exploits to evaluate the exploitability of vulnerabilities \cite{AEG1,AEG2,AEG3,AEG4}. Avgerinos et al. \cite{thanassis2011aeg} proposed an automatic vulnerability mining and exploitation approach, which uses program verification to find the input that can be used and make the program enter an unsafe state (such as Out-of-bounds write and malicious format string). Padaryan et al. \cite{padaryan2015automated} presented a framework, which does not require debug information and could be applied to binary programs, based on program dynamic analysis and symbol execution to construct exploits for stack buffer overflow vulnerabilities. Wu et al. \cite{wu2018fuze} proposed an automated exploitation framework for kernel \emph{Use-After-Free} (UAF) vulnerabilities. They leveraged fuzzing to provide more kernel crashes on contextual environments as a basis for vulnerability exploitation, and then used symbolic execution to exploit the target vulnerability in different contextual environments. In this paper, based on our constructed DA-CG, \emph{GCMiner} dynamically binds exploitable overridden methods to generate valid injection objects for automatic verification.

\section{Conclusion}\label{Conclusion}
Java deserialization vulnerability receives little attention in the academic community despite its severe impact in practice. In this paper, we call for attention to this problem and performs an empirical study to investigate the characteristics of Java deserialization vulnerabilities from the perspective of gadget chain exploitation. Based on our empirical findings, we propose \emph{GCMiner}, a novel gadget chain mining approach which analyzes both explicit and implicit methods calls to identify more exploitable gadget chains and generates valid injection objects through dynamic binding for fuzzing. The evaluation results show that \emph{GCMiner} significantly outperforms state-of-the-art solutions, and discovers 56 unique gadget chains that cannot be identified by baselines.

In the future, we plan to apply \emph{GCMiner} to the industrial scenario to perform a large-scale case study for evaluation. In addition, we also plan to investigate automatic exploit generation techniques for vulnerability reproduction and confirmation.

\section*{Acknowledgment}

This work is supported by the National Natural Science Foundation of China (No.61972335, No. 62202414, No.62002309, No. 62272400); the CCF-AFSG Research Fund (No.CCF-AFSG RF20210022); the Six Talent Peaks Project in Jiangsu Province (No. RJFW-053), the Jiangsu ``333'' Project; the Open Funds of State Key Laboratory for Novel Software Technology of Nanjing University (No.KFKT2022B17), Yangzhou University Top-level Talents Support Program (2019); Xiamen Youth Innovation Fund (3502Z20206036). 

\bibliographystyle{./bibliography/IEEEtran}
\bibliography{./bibliography/IEEEabrv,./bibliography/IEEEexample}

\begin{thebibliography}{10}
\providecommand{\url}[1]{#1}
\csname url@samestyle\endcsname
\providecommand{\newblock}{\relax}
\providecommand{\bibinfo}[2]{#2}
\providecommand{\BIBentrySTDinterwordspacing}{\spaceskip=0pt\relax}
\providecommand{\BIBentryALTinterwordstretchfactor}{4}
\providecommand{\BIBentryALTinterwordspacing}{\spaceskip=\fontdimen2\font plus
\BIBentryALTinterwordstretchfactor\fontdimen3\font minus
  \fontdimen4\font\relax}
\providecommand{\BIBforeignlanguage}[2]{{%
\expandafter\ifx\csname l@#1\endcsname\relax
\typeout{** WARNING: IEEEtran.bst: No hyphenation pattern has been}%
\typeout{** loaded for the language `#1'. Using the pattern for}%
\typeout{** the default language instead.}%
\else
\language=\csname l@#1\endcsname
\fi
#2}}
\providecommand{\BIBdecl}{\relax}
\BIBdecl

\bibitem{DBLP:journals/toplas/HerlihyL82}
M.~Herlihy and B.~Liskov, ``A value transmission method for abstract data
  types,'' \emph{{ACM} Trans. Program. Lang. Syst.}, vol.~4, no.~4, pp.
  527--551, 1982.

\bibitem{SALSA}
J.~C.~S. Santos, R.~A. Jones, C.~Ashiogwu, and M.~Mirakhorli,
  ``Serialization-aware call graph construction,'' in \emph{Proceedings of the
  10th {ACM} {SIGPLAN} International Workshop on the State Of the Art in
  Program Analysis (SOAP)}.\hskip 1em plus 0.5em minus 0.4em\relax {ACM}, 2021,
  pp. 37--42.

\bibitem{DBLP:journals/smr/WeiSBCXL21}
Y.~Wei, X.~Sun, L.~Bo, S.~Cao, X.~Xia, and B.~Li, ``A comprehensive study on
  security bug characteristics,'' \emph{J. Softw. Evol. Process.}, vol.~33,
  no.~10, 2021.

\bibitem{DBLP:journals/chinaf/SunPZLC19}
X.~Sun, X.~Peng, K.~Zhang, Y.~Liu, and Y.~Cai, ``How security bugs are fixed
  and what can be improved: an empirical study with mozilla,'' \emph{Sci. China
  Inf. Sci.}, vol.~62, no.~1, pp. 19\,102:1--19\,102:3, 2019.

\bibitem{Svoboda}
Svoboda, ``Exploiting java deserialization for fun and profit,'' 2016.

\bibitem{DBLP:conf/ccs/DahseKH14}
J.~Dahse, N.~Krein, and T.~Holz, ``Code reuse attacks in {PHP:} automated {POP}
  chain generation,'' in \emph{Proceedings of the 21th {ACM} {SIGSAC}
  Conference on Computer and Communications Security (CCS)}.\hskip 1em plus
  0.5em minus 0.4em\relax {ACM}, 2014, pp. 42--53.

\bibitem{Spring4Shell}
{Spring4Shell}, 2022,
  \url{https://www.picussecurity.com/resource/spring4shell-spring-core-remote-code-execution-vulnerability}.

\bibitem{Spring}
{Spring}, 2022, \url{https://spring.io}.

\bibitem{OWASP}
{OWASP Top Ten 2017 A8: Insecure Deserialization}, 2022,
  \url{https://owasp.org/www-project-top-ten/2017/A8_2017-Insecure_Deserialization}.

\bibitem{Inspector}
I.~Haken, ``Automated discovery of deserialization gadget chains,'' in
  \emph{Proceedings of the Black Hat USA}, 2018.

\bibitem{Serhybrid}
S.~Rasheed and J.~Dietrich, ``A hybrid analysis to detect java serialisation
  vulnerabilities,'' in \emph{Proceedings of the 35th {IEEE/ACM} International
  Conference on Automated Software Engineering (ASE)}.\hskip 1em plus 0.5em
  minus 0.4em\relax {IEEE}, 2020, pp. 1209--1213.

\bibitem{DBLP:journals/ese/LuoPPBPMBHM22}
L.~Luo, F.~Pauck, G.~Piskachev, M.~Benz, I.~Pashchenko, M.~Mory, E.~Bodden,
  B.~Hermann, and F.~Massacci, ``Taintbench: Automatic real-world malware
  benchmarking of android taint analyses,'' \emph{Empir. Softw. Eng.}, vol.~27,
  no.~1, p.~16, 2022.

\bibitem{DBLP:conf/icse/BenzKLBBZ20}
M.~Benz, E.~K. Kristensen, L.~Luo, N.~P. Borges, E.~Bodden, and A.~Zeller,
  ``Heaps'n leaks: how heap snapshots improve android taint analysis,'' in
  \emph{Proceedings of the 42nd International Conference on Software
  Engineering (ICSE)}.\hskip 1em plus 0.5em minus 0.4em\relax {ACM}, 2020, pp.
  1061--1072.

\bibitem{fuzz}
V.~J.~M. Man{\`{e}}s, H.~Han, C.~Han, S.~K. Cha, M.~Egele, E.~J. Schwartz, and
  M.~Woo, ``The art, science, and engineering of fuzzing: {A} survey,''
  \emph{{IEEE} Trans. Software Eng.}, vol.~47, no.~11, pp. 2312--2331, 2021.

\bibitem{fuzz1}
M.~B{\"{o}}hme, C.~Cadar, and A.~Roychoudhury, ``Fuzzing: Challenges and
  reflections,'' \emph{{IEEE} Softw.}, vol.~38, no.~3, pp. 79--86, 2021.

\bibitem{DBLP:conf/pldi/Ruf95}
E.~Ruf, ``Context-insensitive alias analysis reconsidered,'' in
  \emph{Proceedings of the 16th {ACM} {SIGPLAN} International Conference on
  Programming Language Design and Implementation (PLDI)}.\hskip 1em plus 0.5em
  minus 0.4em\relax {ACM}, 1995, pp. 13--22.

\bibitem{sound2}
M.~Reif, F.~K{\"{u}}bler, M.~Eichberg, D.~Helm, and M.~Mezini, ``Judge:
  identifying, understanding, and evaluating sources of unsoundness in call
  graphs,'' in \emph{Proceedings of the 28th {ACM} {SIGSOFT} International
  Symposium on Software Testing and Analysis ({ISSTA})}.\hskip 1em plus 0.5em
  minus 0.4em\relax {ACM}, 2019, pp. 251--261.

\bibitem{ysoserial}
{YSoSerial}, 2022, \url{https://github.com/frohoff/ysoserial}.

\bibitem{DBLP:journals/jss/GantenbeinJ88}
R.~E. Gantenbein and D.~W. Jones, ``The design and implementation of a dynamic
  binding feature for a high-level language,'' \emph{J. Syst. Softw.}, vol.~8,
  no.~4, pp. 259--273, 1988.

\bibitem{POP}
S.~Esser, ``Utilizing code reuse/rop in php application exploits,'' in
  \emph{Proceedings of the Black Hat USA}, 2010.

\bibitem{XStream}
{XStream}, 2022, \url{https://x-stream.github.io/security.html}.

\bibitem{10.1145/3554732}
I.~Sayar, A.~Bartel, E.~Bodden, and Y.~Le~Traon, ``An in-depth study of java
  deserialization remote-code execution exploits and vulnerabilities,''
  \emph{ACM Trans. Softw. Eng. Methodol.}, 2022.

\bibitem{CG}
B.~G. Ryder, ``Constructing the call graph of a program,'' \emph{{IEEE} Trans.
  Software Eng.}, vol.~5, no.~3, pp. 216--226, 1979.

\bibitem{DBLP:conf/icse/Sui0TF20}
L.~Sui, J.~Dietrich, A.~Tahir, and G.~Fourtounis, ``On the recall of static
  call graph construction in practice,'' in \emph{Proceedings of the 42nd
  International Conference on Software Engineering (ICSE)}.\hskip 1em plus
  0.5em minus 0.4em\relax {ACM}, 2020, pp. 1049--1060.

\bibitem{DBLP:conf/issta/NachtigallSB22}
M.~Nachtigall, M.~Schlichtig, and E.~Bodden, ``A large-scale study of usability
  criteria addressed by static analysis tools,'' in \emph{Proceedings of the
  31st {ACM} {SIGSOFT} International Symposium on Software Testing and Analysis
  (ISSTA)}.\hskip 1em plus 0.5em minus 0.4em\relax {ACM}, 2022, pp. 532--543.

\bibitem{thanassis2011aeg}
T.~Avgerinos, S.~K. Cha, B.~L.~T. Hao, and D.~Brumley, ``{AEG:} automatic
  exploit generation,'' in \emph{Proceedings of the 18th Network and
  Distributed System Security Symposium ({NDSS})}.\hskip 1em plus 0.5em minus
  0.4em\relax The Internet Society, 2011.

\bibitem{AEG1}
S.~K. Cha, T.~Avgerinos, A.~Rebert, and D.~Brumley, ``Unleashing mayhem on
  binary code,'' in \emph{Proceedings of the 33rd {IEEE} Symposium on Security
  and Privacy (SP)}.\hskip 1em plus 0.5em minus 0.4em\relax {IEEE}, 2012, pp.
  380--394.

\bibitem{AEG2}
H.~Hu, Z.~L. Chua, S.~Adrian, P.~Saxena, and Z.~Liang, ``Automatic generation
  of data-oriented exploits,'' in \emph{Proceedings of the 24th {USENIX}
  Security Symposium, (USENIX Security)}.\hskip 1em plus 0.5em minus
  0.4em\relax {USENIX} Association, 2015, pp. 177--192.

\bibitem{NVD}
{NVD}, 2022, \url{https://nvd.nist.gov}.

\bibitem{CVE}
{CVE}, 2022, \url{https://cveform.mitre.org}.

\bibitem{Exploit-DB}
{Exploit-DB}, 2022, \url{https://www.exploit-db.com}.

\bibitem{shcherbakov2021serialdetector}
M.~Shcherbakov and M.~Balliu, ``Serialdetector: Principled and practical
  exploration of object injection vulnerabilities for the web,'' in
  \emph{Proceedings of the 28th Annual Network and Distributed System Security
  Symposium (NDSS)}.\hskip 1em plus 0.5em minus 0.4em\relax The Internet
  Society, 2021.

\bibitem{Munoz}
A.~Muñoz and O.~Mirosh, ``Friday the 13th: Json attacks,'' in
  \emph{Proceedings of the Black Hat USA}, 2017.

\bibitem{RMI}
M.~Sharp and A.~Rountev, ``Static analysis of object references in rmi-based
  java software,'' \emph{{IEEE} Trans. Software Eng.}, vol.~32, no.~9, pp.
  664--681, 2006.

\bibitem{DBLP:conf/issta/FourtounisKS18}
G.~Fourtounis, G.~Kastrinis, and Y.~Smaragdakis, ``Static analysis of java
  dynamic proxies,'' in \emph{Proceedings of the 27th {ACM} {SIGSOFT}
  International Symposium on Software Testing and Analysis (ISSTA)}.\hskip 1em
  plus 0.5em minus 0.4em\relax {ACM}, 2018, pp. 209--220.

\bibitem{CommonsCollections1}
{CommonsCollections1}, 2022,
  \url{https://github.com/frohoff/ysoserial/blob/master/src/main/java/ysoserial/payloads/CommonsCollections1.java}.

\bibitem{CHA}
J.~Dean, D.~Grove, and C.~Chambers, ``Optimization of object-oriented programs
  using static class hierarchy analysis,'' in \emph{Proceedings of the 9th
  European Conference on Object-Oriented Programming ({ECOOP})}.\hskip 1em plus
  0.5em minus 0.4em\relax Springer, 1995, pp. 77--101.

\bibitem{Cypher}
R.~Angles, ``A comparison of current graph database models,'' in
  \emph{Proceedings of the 28th International Conference on Data Engineering
  (ICDE)}.\hskip 1em plus 0.5em minus 0.4em\relax {IEEE}, 2012, pp. 171--177.

\bibitem{DBLP:conf/sigsoft/Cheng0BW22}
X.~Cheng, X.~Sun, L.~Bo, and Y.~Wei, ``{KVS:} a tool for knowledge-driven
  vulnerability searching,'' in \emph{Proceedings of the 30th {ACM} Joint
  European Software Engineering Conference and Symposium on the Foundations of
  Software Engineering ({ESEC/FSE})}.\hskip 1em plus 0.5em minus 0.4em\relax
  {ACM}, 2022, pp. 1731--1735.

\bibitem{reflection1}
Y.~Li, T.~Tan, and J.~Xue, ``Understanding and analyzing java reflection,''
  \emph{{ACM} Trans. Softw. Eng. Methodol.}, vol.~28, no.~2, pp. 7:1--7:50,
  2019.

\bibitem{ref1}
B.~Foote and R.~E. Johnson, ``Reflective facilities in smalltalk-80,'' in
  \emph{Proceedings of the 4th {ACM} {SIGPLAN} Conference on Object-Oriented
  Programming Systems, Languages {\&} Applications {(OOPSLA})}.\hskip 1em plus
  0.5em minus 0.4em\relax {ACM}, 1989, pp. 327--335.

\bibitem{ref2}
B.~C. Smith, ``Reflection and semantics in lisp,'' in \emph{Proceedings of the
  11th Annual {ACM} Symposium on Principles of Programming Languages
  (POPL)}.\hskip 1em plus 0.5em minus 0.4em\relax {ACM}, 1984, pp. 23--35.

\bibitem{JQF}
R.~Padhye, C.~Lemieux, and K.~Sen, ``{JQF:} coverage-guided property-based
  testing in java,'' in \emph{Proceedings of the 28th {ACM} {SIGSOFT}
  International Symposium on Software Testing and Analysis (ISSTA)}.\hskip 1em
  plus 0.5em minus 0.4em\relax {ACM}, 2019, pp. 398--401.

\bibitem{Zest}
R.~Padhye, C.~Lemieux, K.~Sen, M.~Papadakis, and Y.~L. Traon, ``Semantic
  fuzzing with zest,'' in \emph{Proceedings of the 28th {ACM} {SIGSOFT}
  International Symposium on Software Testing and Analysis (ISSTA)}.\hskip 1em
  plus 0.5em minus 0.4em\relax {ACM}, 2019, pp. 329--340.

\bibitem{DBLP:journals/ese/ZhouBWSZLZC22}
Z.~Zhou, L.~Bo, X.~Wu, X.~Sun, T.~Zhang, B.~Li, J.~Zhang, and S.~Cao, ``{SPVF:}
  security property assisted vulnerability fixing via attention-based models,''
  \emph{Empir. Softw. Eng.}, vol.~27, no.~7, p. 171, 2022.

\bibitem{Tabby}
X.~Chen, B.~Wang, Z.~Jin, Y.~Feng, X.~Li, X.~Feng, and Q.~Liu, ``Tabby:
  Automated gadget chain detection for java deserialization vulnerabilities,''
  in \emph{Proceedings of the 53rd Annual IEEE/IFIP International Conference on
  Dependable Systems and Network ({DSN})}.\hskip 1em plus 0.5em minus
  0.4em\relax {IEEE}, 2023.

\bibitem{Soot}
{Soot}, 2022, \url{https://soot-oss.github.io/soot}.

\bibitem{Neo4j}
{Neo4j}, 2022, \url{https://neo4j.com}.

\bibitem{DBLP:conf/ccs/KleesRCW018}
G.~Klees, A.~Ruef, B.~Cooper, S.~Wei, and M.~Hicks, ``Evaluating fuzz
  testing,'' in \emph{Proceedings of the 25th {ACM} {SIGSAC} Conference on
  Computer and Communications Security ({CCS})}.\hskip 1em plus 0.5em minus
  0.4em\relax {ACM}, 2018, pp. 2123--2138.

\bibitem{Groovy1}
{Groovy1}, 2022,
  \url{https://github.com/frohoff/ysoserial/blob/master/src/main/java/ysoserial/payloads/Groovy1.java}.

\bibitem{AspectJWeaver}
{AspectJWeaver}, 2022,
  \url{https://github.com/frohoff/ysoserial/blob/master/src/main/java/ysoserial/payloads/AspectJWeaver.java}.

\bibitem{Clojure}
{Clojure}, 2022,
  \url{https://github.com/frohoff/ysoserial/blob/master/src/main/java/ysoserial/payloads/Clojure.java}.

\bibitem{Jython}
{Jython}, 2022,
  \url{https://github.com/frohoff/ysoserial/blob/master/src/main/java/ysoserial/payloads/Jython1.java}.

\bibitem{FUGIO}
S.~Park, D.~Kim, S.~Jana, and S.~Son, ``{FUGIO}: Automatic exploit generation
  for {PHP} object injection vulnerabilities,'' in \emph{Proceedings of the
  31th {USENIX} Security Symposium ({USENIX} Security)}.\hskip 1em plus 0.5em
  minus 0.4em\relax {USENIX} Association, 2022.

\bibitem{Scanner}
{Java Deserialization Scanner}, 2022,
  \url{https://github.com/federicodotta/Java-Deserialization-Scanner}.

\bibitem{objectMap}
N.~Koutroumpouchos, G.~Lavdanis, E.~Veroni, C.~Ntantogian, and C.~Xenakis,
  ``Objectmap: detecting insecure object deserialization,'' in
  \emph{Proceedings of the 23rd Pan-Hellenic Conference on Informatics,
  ({PCI})}.\hskip 1em plus 0.5em minus 0.4em\relax {ACM}, 2019, pp. 67--72.

\bibitem{DBLP:conf/uss/AzadLN19}
B.~A. Azad, P.~Laperdrix, and N.~Nikiforakis, ``Less is more: Quantifying the
  security benefits of debloating web applications,'' in \emph{Proceedings of
  the 28th {USENIX} Security Symposium ({USENIX} Security)}.\hskip 1em plus
  0.5em minus 0.4em\relax {USENIX} Association, 2019.

\bibitem{DBLP:conf/icse/CaoSBWLT22}
S.~Cao, X.~Sun, L.~Bo, R.~Wu, B.~Li, and C.~Tao, ``{MVD:} memory-related
  vulnerability detection based on flow-sensitive graph neural networks,'' in
  \emph{Proceedings of the 44th {IEEE/ACM} 44th International Conference on
  Software Engineering ({ICSE})}.\hskip 1em plus 0.5em minus 0.4em\relax {ACM},
  2022, pp. 1456--1468.

\bibitem{DBLP:journals/infsof/CaoSBWL21}
S.~Cao, X.~Sun, L.~Bo, Y.~Wei, and B.~Li, ``\emph{BGNN4VD}: Constructing
  bidirectional graph neural-network for vulnerability detection,'' \emph{Inf.
  Softw. Technol.}, vol. 136, p. 106576, 2021.

\bibitem{DBLP:journals/infsof/ZhouSXLC19}
T.~Zhou, X.~Sun, X.~Xia, B.~Li, and X.~Chen, ``Improving defect prediction with
  deep forest,'' \emph{Inf. Softw. Technol.}, vol. 114, pp. 204--216, 2019.

\bibitem{DBLP:journals/iet-sen/SubhanWBSR22}
F.~Subhan, X.~Wu, L.~Bo, X.~Sun, and M.~Rahman, ``A deep learning-based
  approach for software vulnerability detection using code metrics,''
  \emph{{IET} Softw.}, vol.~16, no.~5, pp. 516--526, 2022.

\bibitem{Carettoni}
Carettoni, ``Defending against java deserialization vulnerabilities,'' 2016.

\bibitem{DBLP:conf/raid/CristalliVBL18}
S.~Cristalli, E.~Vignati, D.~Bruschi, and A.~Lanzi, ``Trusted execution path
  for protecting java applications against deserialization of untrusted data,''
  in \emph{Proceedings of the 21st International Symposium on Research in
  Attacks, Intrusions and Defenses (RAID)}.\hskip 1em plus 0.5em minus
  0.4em\relax Springer, 2018, pp. 445--464.

\bibitem{AEG3}
A.~Kiezun, P.~J. Guo, K.~Jayaraman, and M.~D. Ernst, ``Automatic creation of
  {SQL} injection and cross-site scripting attacks,'' in \emph{Proceedings of
  the 31st International Conference on Software Engineering (ICSE)}.\hskip 1em
  plus 0.5em minus 0.4em\relax {IEEE}, 2009, pp. 199--209.

\bibitem{AEG4}
A.~Alhuzali, R.~Gjomemo, B.~Eshete, and V.~N. Venkatakrishnan, ``{NAVEX:}
  precise and scalable exploit generation for dynamic web applications,'' in
  \emph{Proceedings of the 27th {USENIX} Security Symposium ({USENIX}
  Security)}.\hskip 1em plus 0.5em minus 0.4em\relax {USENIX} Association,
  2018, pp. 377--392.

\bibitem{padaryan2015automated}
V.~A. Padaryan, V.~Kaushan, and A.~Fedotov, ``Automated exploit generation for
  stack buffer overflow vulnerabilities,'' \emph{Programming and Computer
  Software}, vol.~41, no.~6, pp. 373--380, 2015.

\bibitem{wu2018fuze}
W.~Wu, Y.~Chen, J.~Xu, X.~Xing, X.~Gong, and W.~Zou, ``{FUZE:} towards
  facilitating exploit generation for kernel use-after-free vulnerabilities,''
  in \emph{Proceedings of the 27th {USENIX} Security Symposium ({USENIX}
  Security)}.\hskip 1em plus 0.5em minus 0.4em\relax {USENIX} Association,
  2018, pp. 781--797.

\end{thebibliography}

\end{document}